\PassOptionsToPackage{usenames,dvipsnames,svgnames}{xcolor}
\documentclass[sigplan,10pt]{acmart}

\usepackage{graphicx}
\usepackage{dcolumn}
\usepackage{bm}
\usepackage{times}
\usepackage{outlines}
\usepackage{breakurl}
\usepackage{amsthm}
\usepackage{amsfonts}
\usepackage{ulem}
\usepackage{xspace}
\usepackage{paralist}
\usepackage{xcolor}
\usepackage{xurl}
\usepackage{enumitem}
\usepackage{outlines}
\usepackage{graphicx}
\usepackage{booktabs}
\usepackage{fnpct}
\usepackage{hyperref}
\usepackage{listings}

\newcommand{\hide}[1] 
{
\ifthenelse{\boolean{false}}{#1}{}
}
\newcommand{\secref}[1]{\S\ref{#1}}

\newcommand{\bluelink}[1]{\href{#1}{\textcolor{teal}{\url{#1}}}}

\newcommand{\eg}{e.g.,\xspace}

\newcommand{\eat}[1]{}
\usepackage{algorithm, algorithmicx}
\usepackage{algpseudocode}
\usepackage{multirow}
\usepackage{subcaption}
\usepackage{pifont}
\usepackage{enumitem}
\usepackage{refcount}
\setlist[itemize]{itemsep=0pt, partopsep=0pt, parsep=1pt, topsep=1pt}
\setlist[enumerate]{itemsep=0pt, partopsep=0pt, parsep=1pt, topsep=1pt}
\newcommand{\sys}{SkyServe\xspace}

\newcommand{\policy}{SpotHedge\xspace}

\usepackage[frozencache,cachedir=.]{minted}
\setminted{
  python3=true,
  fontsize=\footnotesize,
  autogobble,
  xleftmargin=1em,
  numbersep=4pt,
  highlightcolor=lightgray,
}
\setmintedinline{fontsize=\small}
\newmintedfile[pythoncode]{python}{}
\newmintedfile[yamlcode]{yaml}{
    baselinestretch=1.2
}

\usepackage{booktabs} 
\usepackage{siunitx} 
\title{SkyServe: Serving AI Models \\ across Regions and Clouds with Spot Instances}
\begin{document}

\acmYear{2025}\copyrightyear{2025}
\setcopyright{rightsretained}
\acmConference[EuroSys '25]{Twentieth European Conference on Computer Systems}{March 30--April 3, 2025}{Rotterdam, Netherlands}
\acmBooktitle{Twentieth European Conference on Computer Systems (EuroSys '25), March 30--April 3, 2025, Rotterdam, Netherlands}
\acmDOI{10.1145/3689031.3717459}
\acmISBN{979-8-4007-1196-1/25/03}



\author{
Ziming Mao$^{*\dagger}$\;
Tian Xia$^{*\dagger}$\;
Zhanghao Wu$^\dagger$\;
Wei-Lin Chiang$^\dagger$\;
Tyler Griggs$^\dagger$\;
\\
Romil Bhardwaj$^\dagger$\;
Zongheng Yang$^\dagger$\;
Scott Shenker$^{\dagger\diamond}$\;
Ion Stoica$^\dagger$\; 
\\
\emph{$^\dagger$UC Berkeley}
\emph{$^\diamond$ICSI}
}
\thanks{$^*$Equal Contributions}

\renewcommand{\shortauthors}{Mao et al.}
\renewcommand{\shorttitle}{SkyServe: Serving AI Models across Regions and Clouds with Spot Instances}
\renewcommand{\authors}{Ziming Mao, Tian Xia, Zhanghao Wu, Wei-Lin Chiang, Tyler Griggs, Romil Bhardwaj, Zongheng Yang, Scott Shenker, Ion Stoica}

\begin{abstract}

Recent years have witnessed an explosive growth of AI models. The high cost of hosting AI services on GPUs and their demanding service requirements, make it timely and challenging to lower service costs and guarantee service quality. While spot instances have long been offered with a large discount, spot preemptions have discouraged users from using them to host model replicas when serving AI models. 

To address this, we propose a simple yet efficient policy, \policy, that leverages spot replicas across different failure domains (\eg regions and clouds) to ensure availability, lower costs, and high service quality. \policy intelligently spreads spot replicas across different regions and clouds to improve availability and reduce correlated preemptions, overprovisions cheap spot replicas than required as a safeguard against possible preemptions, and dynamically falls back to on-demand replicas when spot replicas become unavailable. We built \sys, a system leveraging \policy to efficiently serve AI models over a mixture of spot and on-demand replicas across regions and clouds. We compared \sys with both research and production systems on real AI workloads: \sys reduces cost by 43\% on average while achieving high resource availability compared to using on-demand replicas. Additionally, \sys improves P50, P90, and P99 latency by 2.3$\times$, 2.1$\times$, 2.1$\times$ on average compared to other research and production systems. 

\begin{figure}[t]
    \centering
    \includegraphics[width=.9\linewidth]{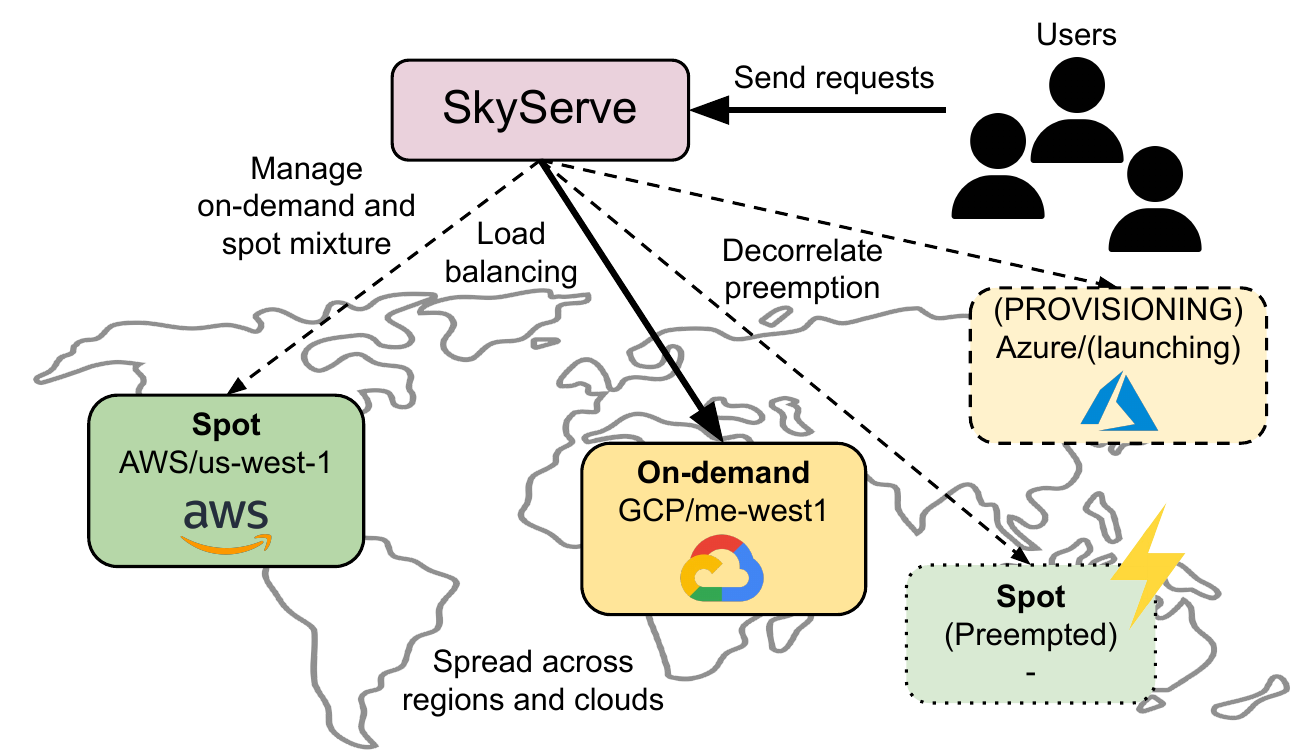}
    \caption{\textbf{\sys Overview.} 
    \sys leverages \policy to intelligently provisions and manages a \textit{mixture of spot and on-demand replicas} across \textit{regions and clouds} to minimize preemptions, improve availability, and reduce cost.}
    \label{fig:skyserve-intro}
\end{figure}

\end{abstract}

\maketitle
\section{Introduction}

Generative AI has experienced explosive growth in the past several years, which has enabled a plethora of new applications, such as large language model (LLM) chatbots~\cite{Google-Bard, openai-chatgpt}, programming assistants~\cite{github-copilot}, image generation~\cite{openai-dalle3}, and writing assistants~\cite{grammarly-ai}. Many companies~\cite{anyscale_endpoints, openai-api, together-ai-endpoint, google-gemini, anthropic-sonnet} offer these models as hosted services on the cloud. A service is composed of multiple model replicas; each replica runs on one or more GPU instances. 
However, serving these models is challenging: not only do they need to be highly available and serve user requests under tight latency constraints, but they are also expensive to operate. Each request served by these models can take several seconds, if not tens of seconds to process.~\cite{SpotServe, seery2024llm, li2023alpaserve}. Compared to traditional web services, AI systems have much higher compute requirements~\cite{openai_compute} and costs~\cite{ai-high-cost}.

There are two reasons for the high cost of serving AI models. First, these models require expensive GPU accelerators. As a result, processing a request can be 10$\times$ more expensive than a traditional search engine query~\cite{ai-high-cost}. Second, real-life AI workloads have frequent and unpredictable traffic spikes (up to 50$\times$ on average~\cite{li2023alpaserve}) and service latency fluctuations~\cite{openai-status}. This results in organizations over-provisioning (provisioning more than required by traffic) more replicas than needed to serve average user traffic or sometimes even provisioning for the peak load, both of which exacerbate the high serving cost. 

\emph{Spot instances} have long been offered by cloud providers as a cost-saving option (8-50\% cost of on-demand; Table \ref{tab:spot-saving}). However, serving AI models on spot GPU instances (or "spot replicas") has a few key challenges. 
First, spot instances can be preempted by the provider at any time, and preemptions of spot GPUs are much more common than spot CPUs (\secref{spot-cpu-fail}). 
Second, when preemptions happen, service quality can degrade due to fewer replicas responding to requests. Naively placing spot replicas in a single region can lead to limited availability and correlated preemptions, where multiple spot instances are preempted simultaneously, potentially resulting in service downtime (\secref{correlated-gpu-preemption}). Finally, recovery of spot replicas can be slow due to the long cold start and provisioning delays, on the order of minutes (\secref{spot-cpu-fail}), or even infeasible, due to the immediate unavailability of spot instances after preemption, in the same zone or region.

Most prior work~\cite{yang2023snape, wagenlander2020spotnik, harlap2017proteus, harlap2018tributary, zhang2019mark, gunasekaran2022cocktail} focuses on the more available spot CPU instances (\secref{spot-cpu-fail}) or on training workloads~\cite{thorpe2023bamboo, athlur2022varuna, yang2021scheduling}. However, serving AI models on spot GPUs requires the system to be robust to frequent preemptions, spot unavailability, and significant cold start delays. Spot preemption warnings cannot address the problem, as the time to find available instances, provision, and load models typically exceeds the best-effort preemption warnings (2 minutes on AWS and 30 seconds on GCP and Azure). Spot instances can also be simultaneously unavailable or preempted in practice (\S\ref{ai-serving-challenges}). As such, serving model replicas on spot GPUs while maintaining high service quality has not been
widely considered viable in practice.

We show that leveraging spot replicas in AI model serving is not only feasible, but can ensure high availability, lower costs, and improve service quality.
We propose a simple yet effective policy, \policy, that uses a dynamic mixture of spot and on-demand replicas across regions and clouds to minimize the cost and improve service latency.
First, \policy improves availability and decorrelates preemptions by spreading spot replicas across wider failure domains (regions or clouds), compared to the common practice of launching in the same zone or region~\cite{google-gke, aws-autoscaling-group, zhang2019mark, miao2023spotserve}.
Using more regions and clouds enlarges the search space where spot instances can be provisioned, significantly improving availability and speeding up recovery. We have observed that a single-region deployment of spot replicas is often not viable due to instance unavailability, leading to service downtime (\secref{ai-serving-challenges}). Second, to avoid degraded service on preemptions, and to ensure high availability, \policy proactively uses on-demand replicas (model replicas running on on-demand instances) as a \textit{fallback}. On-demand instances are typically available if we search and provision across regions and clouds.
Third, \policy mitigates preemption by over-provisioning with cheap spot replicas instead of expensive on-demand replicas. Even when some spot replicas are preempted, these additional over-provisioned replicas will mitigate the impact of preemption. Thus, a service is backed by a dynamic mixture of spot and on-demand replicas.

We implemented \policy in \sys, a real system that provides a unified interface to launch services on a mixture of spot and on-demand replicas across regions and clouds (Figure~\ref{fig:skyserve-intro}).
Users leverage any existing model inference server (\eg  vLLM~\cite{kwon2023efficient}, TGI~\cite{tgi}, Triton~\cite{nvidia-triton}) containing logic to invoke models, and \sys intelligently provisions, maintains, and load-balances a mixture of spot and on-demand replicas across regions and clouds. \sys is compatible with existing state-of-the-art model-level optimizations. 

To evaluate \policy, we deploy \sys on the cloud to serve AI models with spot replicas and experience real-time preemptions. Compared to other research and production systems, \sys improves P50, P90, P99 latency by 2.3$\times$, 2.1$\times$, 2.1$\times$ on average respectively (\secref{end-to-end-evaluation}), and saves cost by 43\% on average compared to using only on-demand replicas while achieving high availability. Additionally, we compare \policy with other policies by replaying real spot traces from AWS and GCP~\cite{wu2024can}. Both experiments show that, with \policy, using spot replicas is feasible in serving AI models and can significantly lower costs and improve service quality. In summary, this paper makes four main contributions:

\begin{itemize}[leftmargin=*]
  \item The design of \policy, a simple yet effective policy 
  that manages a mixture of spot and on-demand replicas across regions and clouds. \policy achieves high availability while improving both cost and service quality. 
  \item The implementation of \sys as a distributed, multi-cloud serving system with mechanisms to scale across spot and on-demand replicas to efficiently serve AI models.
  \item An extensive evaluation of \policy, comparing it to both research and production systems as well as state-of-the-art policies on spot GPU instances. 
  \item An open-source serving system \sys\footnote{We open-sources \sys in \bluelink{https://github.com/skypilot-org/skypilot}.} to facilitate further research and policy design for serving on spot instances.
\end{itemize}

\section{Background and challenges}
\subsection{Serving AI Models on spot instances}
\label{ai-workloads-background}

\begin{figure}[t]
    \centering
    \includegraphics[width=.8\linewidth]{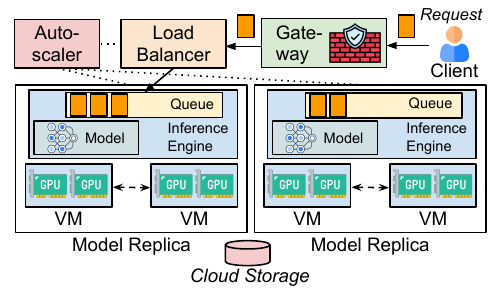}
    \caption{\textbf{An AI service comprises of multiple model replicas; each replica is hosted on one or multiple instances. } Each replica can independently serve user requests without communicating with other replicas.}
    \label{fig:instance-service-replica-figure}
\end{figure}

\paragraph{Serving Systems.} In practice, a \textit{service} (Figure~\ref{fig:instance-service-replica-figure}) hosting AI models comprises of one or multiple model replicas. Each request to the service is load-balanced and routed to one replica. Each replica exposes the model endpoint with an inference engine (\eg vLLM~\cite{kwon2023efficient}, TGI~\cite{tgi}, Triton~\cite{nvidia-triton}) containing logic to invoke models. Within a replica, a model can be partitioned over multiple GPU instances or run on a single GPU instance; there is little traffic across replicas as these replicas can independently serve user requests. We call model replicas running on spot instances \textit{spot replicas}, and model replicas running on on-demand instances \textit{on-demand replicas}. These serving systems are deployed for a wide range of use cases, including chatbots~\cite{openai-chatgpt}, figure generation~\cite{midjourney}, retrieval-augmented generation (RAG)~\cite{lewis2021retrievalaugmentedgenerationknowledgeintensivenlp}, and agentic systems~\cite{durante2024agentaisurveyinghorizons}.

\paragraph{Cost Savings.} AI serving on the cloud is costly~\cite{patel2023inference}. The rise in popularity of AI models~\cite{openai-chatgpt} demands many GPUs to host them. However, GPU instances are expensive~\cite{gpu-high-cost, ai-high-cost}. To compare, the cost of spot instances can be 8\%\text{--}50\% that of on-demand instances (Table~\ref{tab:spot-saving}, \cite{qu2016reliable}), presenting opportunities to reduce the cost of AI serving workloads using spot GPU instances. The cost of spot instances is generally stable over time, though there could be cost differences across zones and regions~\cite{yang2023skypilot}. Despite significant cost savings, the industry has not seen adoption of spot instances in serving AI models, particularly due to spot instance preemptions and unavailability. 

\paragraph{Existing Systems.} Existing systems~\cite{miao2023spotserve, ray-serve, zhang2019mark, harlap2018tributary, aws-autoscaling-group} made promising progress toward reducing cost with spot instances. First, SpotServe~\cite{miao2023spotserve} is a system that adjusts (data, tensor, pipeline) parallelism upon preemption within a replica. However, SpotServe does not consider or implement instance provisioning, placement, or scheduling~\cite{SpotServe}. Second, prior work has investigated training on spot instances~\cite{shang2023spotdnn, harlap2017proteus, thorpe2023bamboo, athlur2022varuna, wagenlander2020spotnik, yang2021scheduling}. However, training aims to finish jobs within deadlines and can be paused and resumed from checkpoints upon spot instance preemption. Hence, training presents significantly different goals from serving. Third, while serverless systems share a similar goal of reducing cost, serverless systems typically execute short-lived tasks and are not suited for long-running model serving; AWS Lambda does not yet support GPUs. Next, we show why these systems are still limited in serving AI models on spot instances.  

\begin{table}[t]
    \centering
\begin{tabular}{lrrrr}
\toprule
     & A100 & V100  & T4  & K80 \\ \midrule
\text{AWS}   & 10\%  & 8-25\%  & 13-17\% & 13-25\% \\
\text{Azure} & 50\%  & 25\%   & 10\% & 10\% \\
\text{GCP}  & 33\%  & 33\% & 14-20\%  & 10\% \\ \bottomrule
\end{tabular}
\vspace{1em}
\caption{Cost of spot GPU instances, in percentage of on-demand cost. Prices obtained via cloud APIs~\cite{aws-pricing,gcp-pricing,azure-pricing} at time of writing (Oct. 23, 2024).
}
\label{tab:spot-saving}
\end{table}

\subsection{Existing single-region systems suffer from limited spot GPU  availability. }
\label{ai-serving-challenges}

Existing systems suffer from limited availability as they primarily focus on single-region deployments.

\begin{figure*}[t]
\centering
\begin{subfigure}{.3\linewidth}
    \centering
    \includegraphics[width=\linewidth]{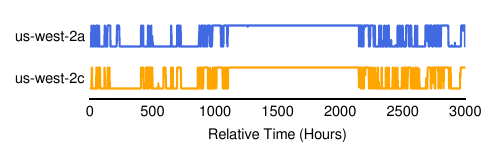}
    \caption{\textbf{Same region} 
    }
    \label{fig:correlation-single-region}
\end{subfigure}
~
\begin{subfigure}{.32\linewidth}
    \centering
    \includegraphics[width=\linewidth]{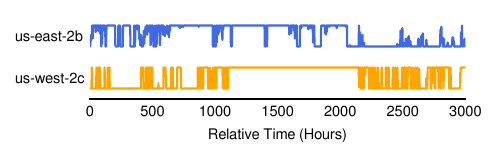}
    \caption{\textbf{Different regions}
    }
\label{fig:correlation-multiple-region}
\end{subfigure}
~
\begin{subfigure}{.36\linewidth}
    \centering    \includegraphics[width=\linewidth]{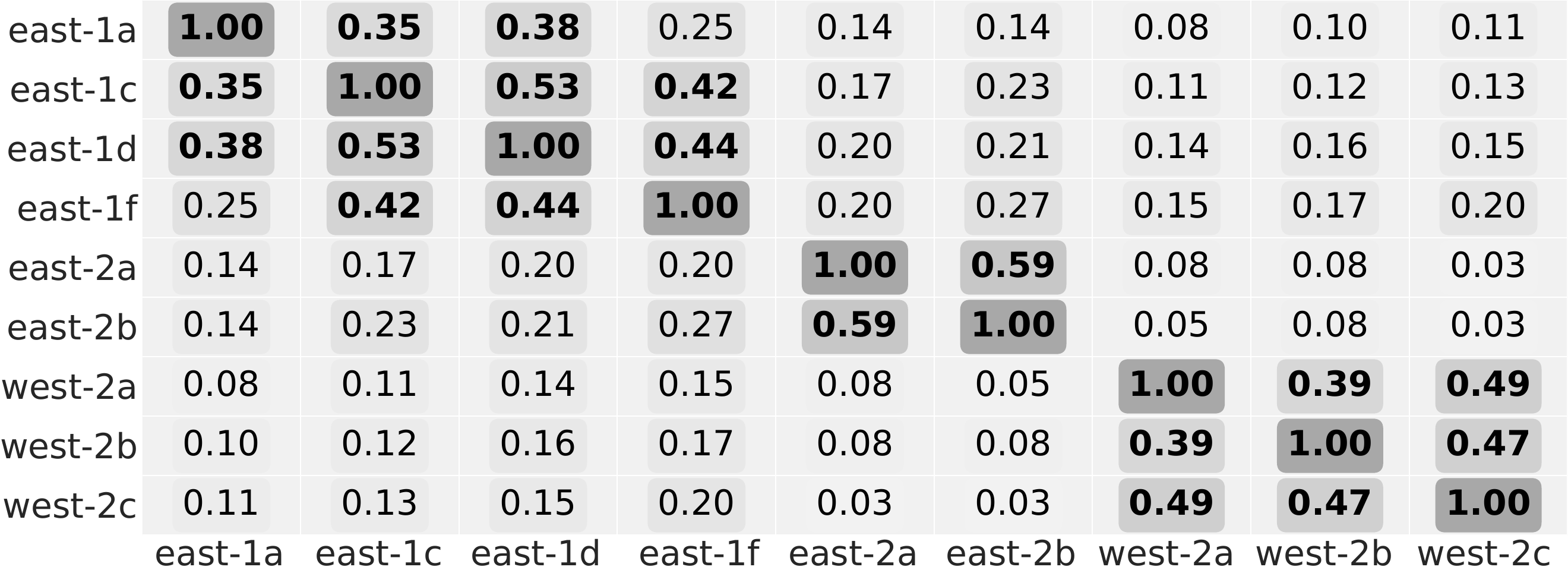}
    \caption{\textbf{Correlation} 
    } 
    \label{fig:preemption-heatmap}
\end{subfigure}
\caption{\textbf{(a)} Correlated spot GPUs preemptions within the same region; \textbf{(b)} Lack of correlation across regions. Both (a) and (b) are from a 2-week trace (\secref{microbenchmark}) for 4 p3.2xlarge (V100) instances. To collect the trace, we try to maintain the desired number of spot instances, record preemption, and replenish any preempted instances. Each vertical line indicates either preemption (from higher to lower) or a successful launch (from lower to higher). \textbf{(c)} Correlated preemptions across 8 zones in 3 regions on AWS for V100 GPU. Each cell shows the correlation between two zones indicated by the row and column labels. The values are Pearson Correlation (with $p < 0.01$), and we bold correlation $>= 0.3$. Intra-region has more correlation among \{east-1a, east-1c, east-1d, east-1f\}, \{east-2a, east-2b\}, \{west-2a, west-2b\} whereas there is little to no inter-region correlation.} 
\label{fig:correlated-preemptions}
\end{figure*}

\paragraph{Unavailability of spot GPUs in one region.} 

GPUs often experience shortages~\cite{companies-gpu-shortage} within a region. We have observed spot GPU unavailability across zones of the same region, either due to the region running out of capacity, or quota issues. For example, in one trace we analyzed (AWS 2, \secref{microbenchmark}), 33.1\% of time spot GPUs experience unavailability across all zones in a single region. In our end-to-end evaluation (\secref{end-to-end-evaluation}), region us-west-2 has experienced unavailability for 21\% of time. This observation generalizes across multiple GPU instance types and is one of the key limitations of prior work: the service can be unavailable if spot GPU instances are simultaneously unavailable in a single region. A single-region system will experience service disruptions when another spot GPU instance cannot be provisioned in the zone or region experiencing preemptions. This significantly limits the practical deployment of AI models on spot instances: in our evaluation with pure spot deployment of AWS Autoscaling Group (\secref{end-to-end-evaluation}), 49.4\% of requests experience failures or time out, either due to spot instance unavailability or limited spot capacity to serve the full load. SpotServe, while being able to adjust parallelization strategies upon preemption, has a failure rate of 2.0\text{--}75.9\% depending on the region it is deployed. Preemptions are rarely isolated and independent; a region experiencing spot instance preemption is likely to continue facing spot instance preemption for some time. Serving AI models on spot instances requires the system to \textit{maintain} the desired number of instances to sustain the full load, in addition to being able to recover from isolated preemptions.

\begin{figure*}[t]
\centering
\begin{subfigure}{.35\linewidth}
    \centering    \includegraphics[width=\linewidth]{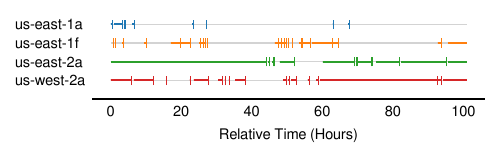}
    \caption{\textbf{Availability of spot GPUs} 
    } 
    \label{fig:volatility-gpu}
\end{subfigure}
~
\begin{subfigure}{.35\linewidth}
    \centering
    \includegraphics[width=\linewidth]{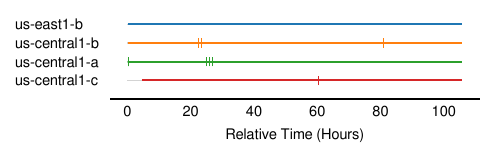}
    \caption{\textbf{Availability of spot CPUs}}
    \label{fig:volatility-cpu}
\end{subfigure}
\caption{Spot GPUs (p3.2xlarge) experience more preemptions than spot CPUs (c3-highcpu-176). Horizontal lines represent the available period. Vertical bars are changes from available to unavailable, followed by grey gaps indicating the unavailable period.}
\end{figure*}

\paragraph{Correlated spot GPU preemptions within a region.} 
\label{correlated-gpu-preemption}
During periods where spot instances are available in a region, we observe correlated preemptions for zones in the same region (Figure~\ref{fig:correlation-single-region}). 
To compare, there is much less correlation for spot GPUs across different regions (Figure~\ref{fig:correlation-multiple-region}). This trend is corroborated by analyzing a 2-month, 8-zone trace and observing correlation 
across zones of the same region, and little correlation across different regions (Figure~\ref{fig:preemption-heatmap}), showing that the trend is general across different regions. This observation is recently mentioned by another study done by CockroachDB~\cite{taylor2020spotinstances} that similarly observed correlated preemptions within zones and regions. 
This means that if the majority of the replicas are hosted on spot instances in a single region, simultaneous preemptions could result in service losing significant capacity and even in service outage before the system can respond in time. From a 3-week trace of 16 p3.2xlarge instances across three zones on AWS, we observe that from the first spot instance preemption, 83--97\% of the time a preemption occurs in a zone, at least one more will follow within 5 minutes. From a 3-day trace of 4 a2-ultragpu-4g instances in 6 zones on GCP, we observe that 34-95\% of time other spot instances of the same zone are preempted within 150 seconds. Since we target serving workload, even a small likelihood of simultaneous preemption cannot be overlooked. Unfortunately, none of the prior work considers single-region unavailability or correlated preemptions.

\subsection{Existing spot CPU-focused policies failed to work on spot GPUs. }
\label{spot-cpu-fail} \label{spot-gpu-different-cpu}

The majority of the systems~\cite{yang2023snape, wagenlander2020spotnik, harlap2017proteus, harlap2018tributary, zhang2019mark, gunasekaran2022cocktail} targeting spot instances use CPU instances. In particular, these systems use preemption warnings to mitigate service disruptions. 
We categorize the limitations of those systems as follows.

\paragraph{Higher preemption rate. }

AWS Spot Instance Advisor~\cite{aws-spot-instance-advisor} shows that the probability of spot GPUs being interrupted (>20\%) is typically much higher than spot CPUs (<5\%). Figure~\ref{fig:volatility-gpu} and Figure~\ref{fig:volatility-cpu} compare two spot obtainability traces 
(\secref{experiment-methodology}) 
between spot GPU and CPU instances. Spot GPUs (16.7\%\text{--}90.4\% available) are more volatile and unobtainable~\cite{lee2017deepspotcloud}, whereas spot CPUs (95.6\%\text{--}99.9\% available) 
experience fewer preemptions. While systems that use spot CPUs might be fine just waiting and recovering from infrequent preemptions, serving with high-demand spot GPUs requires the system to be robust to frequent preemptions and potential unavailability.

\paragraph{Longer time to provision and deploy a new replica.}

Many CPU targeting systems use a simple policy to keep the service running: launch a new replica upon receiving a preemption warning.
This is based on the assumption that the time to deploy a replica, including provisioning the instance(s) and loading the model into GPU, is shorter than the preemption warning (2 minutes for AWS~\cite{amazon2015spot} and 30 seconds for Azure~\cite{microsoft2024spot} and GCP~\cite{google2023spot}).
Therefore, the backup instances can become ready before the old instance gets preempted, resulting in little to no service downtime. Does preemption warning minimize the disruption caused by preemption? 

Unfortunately, spot GPU instances can take a long time to provision due to potential unavailability and large AI model endpoints taking longer to be ready. Previous studies~\cite{anyscale_endpoints, scale_cold_start_time, zhang2019mark, miao2023spotserve} have documented that initiating large AI model endpoints can take several minutes, involving instances provisioning and transferring model weights to the GPU. 
We run a simple experiment and find that the time taken to provision an instance with a pre-installed image and deploy an AI model endpoint (\text{Llama-2-7b} on vLLM~\cite{kwon2023efficient}) is 183s on AWS, already exceeding the 2 minutes preemption warning, even without accounting for the time to find available spot GPUs in the event of unavailability.  
While recent systems, such as ServerlessLLM~\cite{fu2024serverlessllm}, can reduce the time to load the model into the GPU, the time needed to find and provision the instance \textit{in the presence of unavailability} is unlikely to be reduced. Spot preemption warnings are also best-effort~\cite{google2023spot, microsoft2024spot, amazon2015spot}. There is no guarantee that the system will be notified and bring up an instance in time. 
Thus, it is challenging to serve these models on spot GPUs just by relying on preemption warnings to solve the problem. 
If managed naively, preempted spot GPUs can both cause service unavailability, due to the time it takes to replenish a ready-to-serve replica, and higher cost, since users are still billed during the cold start period.

\subsection{Existing systems with static spot-on-demand mixture either are costly or have poor availability.}
\label{static-node-pool}

Several systems support serving on both spot and on-demand replicas~\cite{aws-autoscaling-group, ray-serve, miao2023spotserve} to mitigate preemptions and unavailability. These systems require setting fixed and predefined node pools of either spot or on-demand instances. If spot replicas are preempted, the traffic will be re-distributed over on-demand replicas in the on-demand node pools. 
The mixture size (i.e. node pool size) is static, for example, AWS Autoscaling Group can statically maintain $10\%$ of on-demand replicas at all times. This allows on-demand replicas to serve as the base capacity for the service in the event of preemption.

\paragraph{Using a fixed pool of on-demand replicas is unnecessary and costly when spot replicas are available.} During periods with high spot obtainability, systems with static pools of on-demand and spot replicas still keep the costly on-demand replicas, instead of leveraging more spot replicas. 
For example, we run AWS Autoscaling Group~\cite{aws-autoscaling-group} (ASG) with on-demand node pool of size 1 and spot node pool of size 4 for g5.48xlarge instances. During periods where spot instances are available, ASG maintains one on-demand replica throughout, providing base service capacity. Even with a single on-demand replica, this increases the total cost by 1.56$\times$ compared to using a pure spot deployment. The on-demand g5.48xlarge replica cost constitutes 52\% of the total cost, as its hourly cost (\$16.3) is significantly higher than that of spot instance (\$4.9). As such, using always-on on-demand replicas can be expensive and unnecessary when spot instances are available.

\paragraph{A fixed pool of spot instances may fail to provision when spot is unobtainable.} When the system fixes the spot node pool size, it will continue provisioning for the specified node pool size even when spot instances are unobtainable, rather than launch more on-demand instances to cover for lost capacity in the spot pool. 
As the system retries to fulfill the lost spot capacity, it incurs additional costs due to provisioning and the subsequent quick preemption or lack of capacity (\S\ref{end-to-end-evaluation}). Fixed spot node pools also result in bad service quality. 
For example, we ran the previous ASG deployment during periods with spot volatility. We observe that using such a deployment will result in a request failure rate of 36\%, since for some time intervals the deployment only has one on-demand instance to serve user requests and is severely overloaded. 

Instead of maintaining a static mixture of spot and on-demand replicas, the system should launch more on-demand replicas to dynamically cover the lost spot replica capacity and scale them down when spot instances become more obtainable. The challenge is how to design such a system to maintain a dynamic mixture of spot and on-demand replicas.

\section{SpotHedge}

We propose \policy to address the aforementioned challenges. To overcome limited availability and correlated preemptions in a single region (\S\ref{ai-serving-challenges}), \policy dynamically provisions spot replicas across regions and clouds based on their preemption risk. To lower cost with good availability (\S\ref{static-node-pool}), \policy adaptively maintains spot and on-demand mixture. We discuss our designs below: 

\subsection{Placing spot replicas dynamically across regions and clouds}
\label{Spot-multi-region}

\label{Spot-placement}

\policy addresses the challenges of limited spot obtainability in a single region by using spot instances across different regions and clouds. Before discussing how \policy selects and provisions spot instances, we discuss and compare prior single-region multi-zone policies in other systems.

\paragraph{Comparing alternative policies.} 

Assume $N$ zones and for each zone $i$, 
preemptions follow Poisson distribution with rate parameter $\lambda_i$. Spot instance lifetime is $\frac{1}{\lambda_i}$. 
The average number of preemptions $\mathbb{E}[K_i]$ in zone $z_i$ over an observation window $T$ is $\mathbb{E}[K_i] = T\lambda_i$, assuming that spot instance lifetime is much greater than the cold start delay $d$.

\textit{Static Spread Policy} (used by AWS Autoscaling Group~\cite{aws-autoscaling-group} and MArk~\cite{zhang2019mark} in a single region): Consider a simple policy where we maintain an even \textit{static} spread of $n$ spot instances to $N$ zones, such that each zone is given $\frac{n}{N}$ number of replicas. The expected total number of preemptions $K$ over $T$ is: $\mathbb{E}[K] = nT \frac{1}{N}\sum_{i=1}^N \lambda_i$. $\mathbb{E}[K]$ will be dominated by highly-preempting zones with large $\lambda_i$. This is not ideal; intuitively, when a zone experiences more preemptions, an intelligent policy should avoid provisioning more spot instances in that zone.

\textit{Round Robin} (used by Ray Serve~\cite{ray-serve}, and GKE~\cite{aws-gke-od-spot-mix} in a single region): Round Robin policy can be used to mitigate the above issue. When a spot instance gets preempted in a zone $i$, it gets relaunched in the next zone in the same region.
For a long-running service, the average spot lifetime is $\frac{1}{N}\sum_{i=1}^N\frac{1}{\lambda_i}$ over $N$ zones. The expected total number of preemptions is: $\mathbb{E}[K] = nT(\frac{N}{\sum_{i=1}^N\frac{1}{\lambda_i}})$. 
Since $\frac{1}{N}\sum_{i=1}^N \lambda_i$ is larger than $\frac{N}{\sum_{i=1}^N\frac{1}{\lambda_i}}$, the Round Robin strategy will lead to fewer preemptions. However, Round Robin is not optimal as it does not remember the highly-preempting zones: it might keep launching instances in them.

\textit{Observation}: If we track $\lambda_i$ of different zones, we can avoid highly-preempting zones and further lower $\mathbb{E}[K]$. Furthermore, a single region typically has a small number of zones with the required GPU. When $\lambda$ is large across these zones, $\mathbb{E}[K]$ will be large regardless of how we place instances. Therefore, we should expand the number of zones from which we launch spot instances by using multiple regions and clouds. 

\paragraph{Dynamic Placement Policy.}
As such, we propose the following spot replica placement policy: \textit{Dynamic Placement} (Algorithm~\ref{alg:dynamic-placement}). To avoid highly-preempting zones, the policy tracks which zones are more likely to experience preemption and dynamically selects zones from available zones. Let $Z$ be all enabled zones with the required instance type that satisfy user requirements, such as avoiding particular zones or regions for regulation or latency constraints. The algorithm keeps two lists: $Z_A$ is a list of available zones initialized to $Z$; $Z_P$ is a list of highly-preempting zones initialized to be empty. 
If a replica is preempted in zone $z$, $z$ is moved to $Z_P$. If a replica is successfully launched and ready 
in zone $z$, $z$ is moved to $Z_A$. Whenever a new replica needs to be launched, it is drawn from $Z_A$, prioritizing zones with fewer current spot placements and zones with lower cost. The policy can additionally probe different zones to maintain $Z_P$ and $Z_A$. 
When there are fewer than two zones left in $Z_A$, \policy triggers zone rebalancing and proactively moves all zones from $Z_P$ to $Z_A$. This prevents us from having only a single zone in $Z_{A}$ and subsequent spot replicas being placed on that single available zone, risking simultaneous preemptions of all spot instances.

\begin{figure}[t]
    \centering
    \begin{subfigure}{.47\linewidth}
        \includegraphics[width=\linewidth]{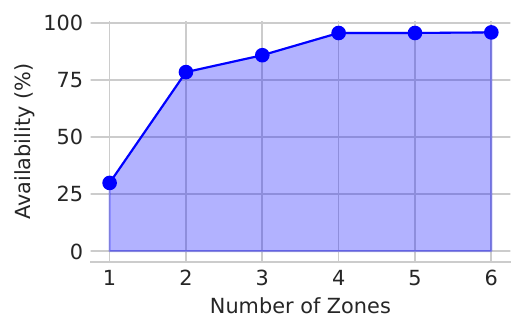}
        \caption{\textbf{A100-80GB:4}}
        \label{fig:avail-vs-zone-a100}
    \end{subfigure}
    ~
    \begin{subfigure}{.47\linewidth}
        \includegraphics[width=\linewidth]{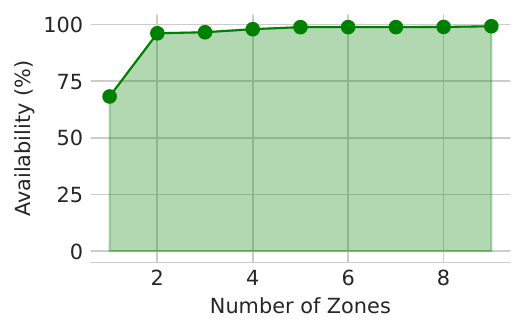}
        \caption{\textbf{V100}} 
        \label{fig:avail-vs-zone-v100}
    \end{subfigure}
    \caption{\textbf{Service availability improves as the number of zones and regions considered increases}. Fig~\ref{fig:avail-vs-zone-a100} uses a 3-day trace for a2-ultragpu-4g in 6 zones and 5 regions. Fig~\ref{fig:avail-vs-zone-v100} uses a 2-month trace for p3.2xlarge in 9 zones and 3 regions. } 
    \label{fig:avail-vs-zone}
\end{figure}

\paragraph{Expanding the spot search space from a single zone to multiple regions and clouds.} To have many zones in $Z_A$, the set of enabled zones has to span multiple regions. To mitigate correlated preemptions, \policy diversifies the set of zones where spot preemptions can occur. This also addresses the challenge of single-region unavailability as discussed in \S\ref{ai-serving-challenges}. As such, \policy extends the search space from a single region to multiple regions. From analyzing traces (Fig.~\ref{fig:avail-vs-zone}), we observe significant improvement to availability (29.9\%$\rightarrow$95.8\% for A100, 68.2\%$\rightarrow$99.2\% for V100) as we increase the search space from a single zone to multiple regions. A preemption-aware placement policy, coupled with an expanded search space across regions and clouds, can significantly improve service availability and quality.

\begin{algorithm}[h]
\caption{Dynamic Placement}\label{alg:dynamic-placement}
\begin{algorithmic}[1]
\State $Z_{A} \gets Z, Z_{P} \gets \varnothing$ \Comment{Initially all zones are available.}

\Procedure{Handle-Preemption}{$z$}: 
    \If{$z \in Z_{A}$}  \Comment{Move zone $z$ to $Z_p$}
    \State{Remove $z$ from $Z_{A}$}
    \State{Add $z$ to $Z_{P}$}
  \EndIf
  \If{Size of $Z_{A}$ $< 2$}
    \State $Z_{A} \gets Z_{A} + Z_{P}, Z_{P} \gets \varnothing$
  \EndIf
\EndProcedure

\Procedure{Handle-Launch}{$z$}: 
    \If{$z \in Z_{P}$}  \Comment{Move zone $z$ to the available list}
    \State{Remove $z$ from $Z_{P}$}
    \State{Add $z$ to $Z_{A}$}
  \EndIf
\EndProcedure

\Procedure{Select-Next-Zone}{$C$}: 
    \State{$Z_A' \gets Z_A \setminus C$}  \Comment{$C$: currently launched zones}
    \If{$Z_{A}'$ is not empty}
        \State{Return \Call{Min-Cost}{$Z_{A}'$}}
    \EndIf
    \State{Return \Call{Min-Cost}{$Z_{A}$}}
\EndProcedure
\end{algorithmic}
\end{algorithm}

\begin{figure}[t]
    \centering
    \begin{subfigure}{.5\linewidth}
        \includegraphics[width=\linewidth]{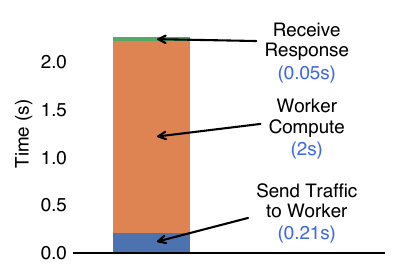}
        \caption{\textbf{Request latency}} 
        \label{fig:request-latency-breakdown}
    \end{subfigure}
    ~
    \begin{subfigure}{.47\linewidth}
        \includegraphics[width=\linewidth]{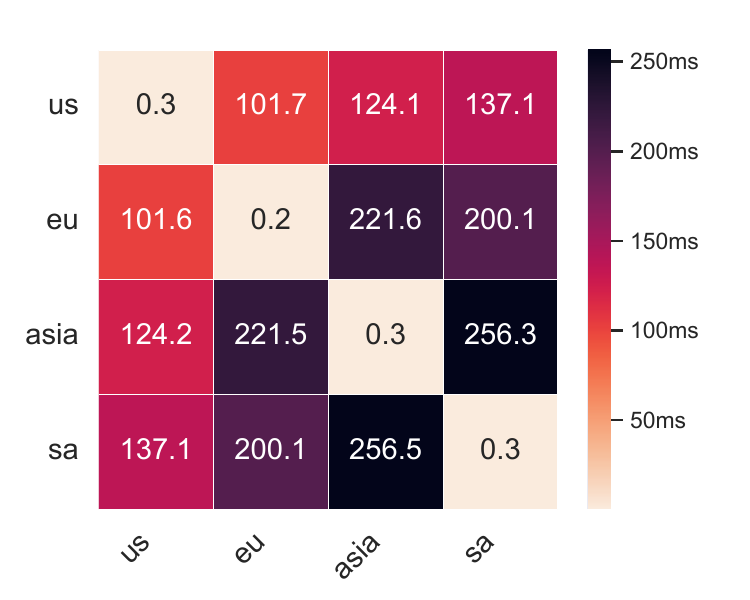}
        \caption{\textbf{Network latency}}
        \label{fig:network-latency}
    \end{subfigure}
    \caption{\textbf{Latency Characteristics of AI Services}. Fig.~\ref{fig:request-latency-breakdown} measures the latency breakdown of a Vicuna-13B endpoint, serving a request with 20 input and 44 output tokens. Fig.~\ref{fig:network-latency} measures round trip network latency between different regions of GCP.}
\end{figure}

\paragraph{Request processing dominates the end-to-end latency.}
One concern from systems practitioners about serving model replicas from other regions is the round-trip latency (\S\ref{ai-workloads-background}). 
We argue that the latency can be improved by serving from a remote region when the spot GPUs in the local region are unavailable or overloaded. For some applications, round-trip latency is much smaller than request processing time and potential queueing delay. 
Processing a single request takes several seconds, if not tens of seconds (Figure~\ref{fig:request-latency-breakdown}, \secref{end-to-end-evaluation}, ~\cite{SpotServe, seery2024llm}) for large AI models even on a local cluster, whereas network latency is much smaller, for example, around 100ms round trip between US and Europe (Figure~\ref{fig:network-latency}). Indeed, OpenAI recently~\cite{morikawa_2023} expanded to multiple regions for availability and matches our observation that geographic placement matters less for large AI model inference since the query latency is much larger than network latency. AI models have been growing in size and computation time and they are expected to do so in the future. With the rise of complex AI systems that use chained API calls~\cite{zheng2023efficiently, chen2024more}, agents, Tree of Thoughts~\cite{openai-chatgpt}, or RAG~\cite{khattab2023dspy}, we expect end-to-end latency to become more critical.
While serving user requests from a different region could increase Time-to-first-token (TTFT)\footnote{The amount of time it takes for a language model to generate the very first token of its response after receiving a prompt.}, \policy benefits from improved availability and better average and tail end-to-end latency. We further discuss this trade-off in \S\ref{discussion-future-work}. 

\subsection{Adjusting spot and on-demand mixture dynamically based on the risk of spot preemption}
\label{Spot-On-Demand-mixture} 

Most existing systems~\cite{ray-serve, google-gke} only use static node pools and don't change the mixture dynamically when preemption happens: launch more spot replicas when spot instances become available, fall back to on-demand replicas when spot market becomes volatile. We have shown in \S\ref{static-node-pool} that static node pools either result in poor availability or high cost, which requires an adaptive spot and on-demand mixture based on estimated risk of spot preemption. To derive the dynamic mixture policy, let $S(z, t)$ be the number of launched spot replicas at time $t$ in zone $z$, and $O(t)$ be the number of launched on-demand replicas at $t$. The total number of launched spot replicas across zones is $S(t) = \sum_{z \in Z} S(z,t)$. Let $N_{Tar}(t)$ be the target number of ready instances decided by the user or an autoscaling policy at $t$ based on the load.

\paragraph{Overprovisioning with $N_{Extra}(t)$.}

\sys overprovisions cheap spot replicas to mitigate preemptions and cold start delay. We use $N_{Extra}(t)$ to denote the number of spot replicas to overprovision at $t$. These spot replicas serve as \textit{buffer} in the event of preemptions. Intuitively, even when some spot replicas are preempted and the system is provisioning new replicas, these \textit{additional} overprovisioned replicas will prevent the rest of the replicas from being overloaded.
Interestingly, we find that a small number of $N_{Extra}(t)$ is often sufficient in practice (Figure~\ref{fig:sensitivity-experiment-aws-2}) if the spot replicas are de-correlated across regions and clouds. We find empirically that these overprovisioned spot replicas constitute only a fraction of the cost, and are cheaper than on-demand replicas (\S\ref{static-node-pool}).

\paragraph{Deciding the number of fallback on-demand replicas.}
Since having fixed on-demand node pools leads to higher costs, we propose a new policy: \textit{Dynamic Fallback}. 
The policy initializes with $N_{Tar}(t) + N_{Extra}(t)$ spot replicas. If a spot replica is preempted, the policy launches an on-demand replica and keeps trying to have $N_{Tar}(t) + N_{Extra}(t)$ spot replicas \textit{at the same time}. Let $S_r(t)$ be the number of ready spot replicas. The policy tries to maintain $O(t)$  (but possibly not all ready) on-demand replicas where 
$O(t) = \min(N_{Tar}(t), N_{Tar}(t)\newline + N_{Extra}(t) - S_r(t))$.
This policy provisions on-demand replicas to replenish the lost spot capacity, where $O(t) \leq N_{Tar}(t)$.

Intuitively, when we have a spot replica preempted, Dynamic Fallback quickly launches an on-demand replica. On-demand replicas serve as a reliable fallback; we have observed in \S\ref{end-to-end-evaluation} that on-demand replicas are typically available across regions. While the on-demand replicas are provisioning, the overprovisioned spot replicas ensure little service disruption. 
When the spot replicas become available, \policy scales down these on-demand replicas and instead serves entirely on spot replicas. The cost of Dynamic Fallback is relatively small as these on-demand replicas will be terminated once we have $N_{Tar}(t) + N_{Extra}(t)$ spot replicas ready. In the event of spot unobtainability, on-demand replicas (up to $N_{Tar}(t) $) are necessary to ensure that the service still meets the availability requirement. 

\subsection{Putting these together}

\label{omniscient}

\policy first decides $N_{Tar}(t)$ and $N_{Extra}(t)$ based on the traffic. Next, it decides the spot-on-demand mixture (i.e. the number of spot replicas and on-demand replicas at $t$, or $S(t)$ and $O(t)$). Lastly, it assigns $S(t)$ spot replicas to different zones, regions, and clouds based on their risk of spot preemption. 

\begin{figure}[t]
    \centering
    \includegraphics[width=.8\linewidth]{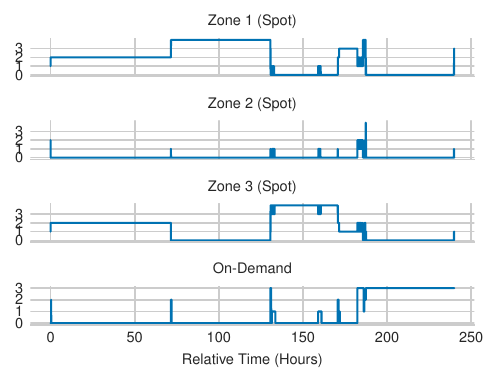}    
    \caption{\textbf{\policy Example}. Launch 4 spot replicas with Dynamic Placement (Alg.~\ref{alg:dynamic-placement}) and Dynamic Fallback. The y-axis refers to the number of launched replicas at a given time.} 
    \label{fig:single-trace}
\end{figure}

\paragraph{\policy illustration.}
We illustrate \policy with an example (Figure~\ref{fig:single-trace}). 4 spot replicas are distributed among three zones (zone 1, 2, 3).  \policy initially fails to launch spot replicas in zone 2, as such, they are launched in zone 1 and zone 3 as zone 2 is moved to $Z_p$. On-demand replicas are simultaneously launched as a fallback but quickly terminated once spot replicas have been launched and are in service. 
When zone 3 becomes unavailable, \policy moves replicas to zone 1. Similarly, when zone 1 becomes unavailable, \policy launches replicas in zone 3. 
Zone 3 later experiences preemptions, prompting \policy to re-try in zones 1 and 2. On-demand fallback is triggered at the end when all zones lose availability. 

\paragraph{Omniscient.}
We propose an Omniscient policy that requires a complete spot obtainability trace (infeasible in practice). With Integer Linear Programming (ILP), we represent the policy as a cost-minimization problem. 
We use  $S_r(z,t)$, $S(z,t)$ to denote the number of (ready) spot replicas launched in zone $z$ and time $t$, and $O_r(t)$, $O(t)$ to denote the number of (ready) on-demand replicas at time $t$. $C(z,t)$ denotes the number of launchable spot replica capacity at zone $z$ and time $t$, typically unknown to users in an online setting. $M(t)$ is a binary variable denoting whether $S_r(t) + O_r(t) \geq N_{Tar}(t)$, recording whether the policy has satisfied the target number of replicas. $d$ is the cold start delay. $k$ is the cost ratio between spot and on-demand replicas. $N_{\max}$ is the maximum required number of replicas. The normalized cost $C$ can be expressed by:
$
C = \sum_{t=0}^T [\sum_{z \in Z} S(z,t) + k O(t)]
$. 
We express a \textit{resource availability} constraint with $Avail_{Tar}$, the percentage of time at least $N_{Tar}$ replicas are ready.
Let $T$ be an interval. 

{
\setlength{\jot}{1pt}
\small
\begin{gather}
    \label{eqn:cost-min}
    \min \sum_{t=0}^T [\sum_{z \in Z} S(z,t) + kO(t)] \\
    \label{eqn:avail-req}
    \sum_{t=0}^T M(t) \geq T \times Avail_{Tar}\\
    \label{eqn:Spot}
    \forall z \in Z, \forall t(0 \leq t \leq T), S(z,t) \leq C(z,t)\\
    \label{eqn:cold-start-cases}
    \forall t(d \leq t \leq T), \forall t'(t - d < t' \leq t) \begin{cases} S(t') \geq S_r(t) \\
    O(t') \geq O_r(t) \\
    \end{cases} \\
    \forall t(0 \leq t \leq T) \begin{cases}
    M(t) \times N_{\max} \geq S_r(t) + O_r(t) - N_{Tar}(t) \\
    (1 - M'(t)) \times N_{\max} \geq N_{Tar}(t) - S_r(t) - O_r(t)
    \end{cases}
    \label{eqn:m(t)}
\end{gather}
\normalsize
}

The Omniscient policy minimizes the cost (Eq. \ref{eqn:cost-min}) while respecting the availability constraint (Eq. \ref{eqn:avail-req}). Eq. \ref{eqn:Spot} limits the number of launchable spot replicas based on the spot capacity. Eq.~\ref{eqn:cold-start-cases} calculates the number of ready replicas given cold start delay $d$. For any time that is up to cold start delay $d$ ago, the launched spot (on-demand) replicas at that time should be greater or equal to the current number of ready spot (on-demand) replicas, because it will take time $d$ for these launched instances to be ready. Eq.~\ref{eqn:m(t)} calculates $M(t)$ to indicate whether $S_r(t) + O_r(t) \geq N_{Tar}(t)$, which is then used to calculate whether the policy meets $Avail_{Tar}$. We compare various policies with the Omniscient optimal policy in \S\ref{microbenchmark}.

\section{System design}
\begin{figure}[t]
    \centering
    \includegraphics[width=.9\linewidth]{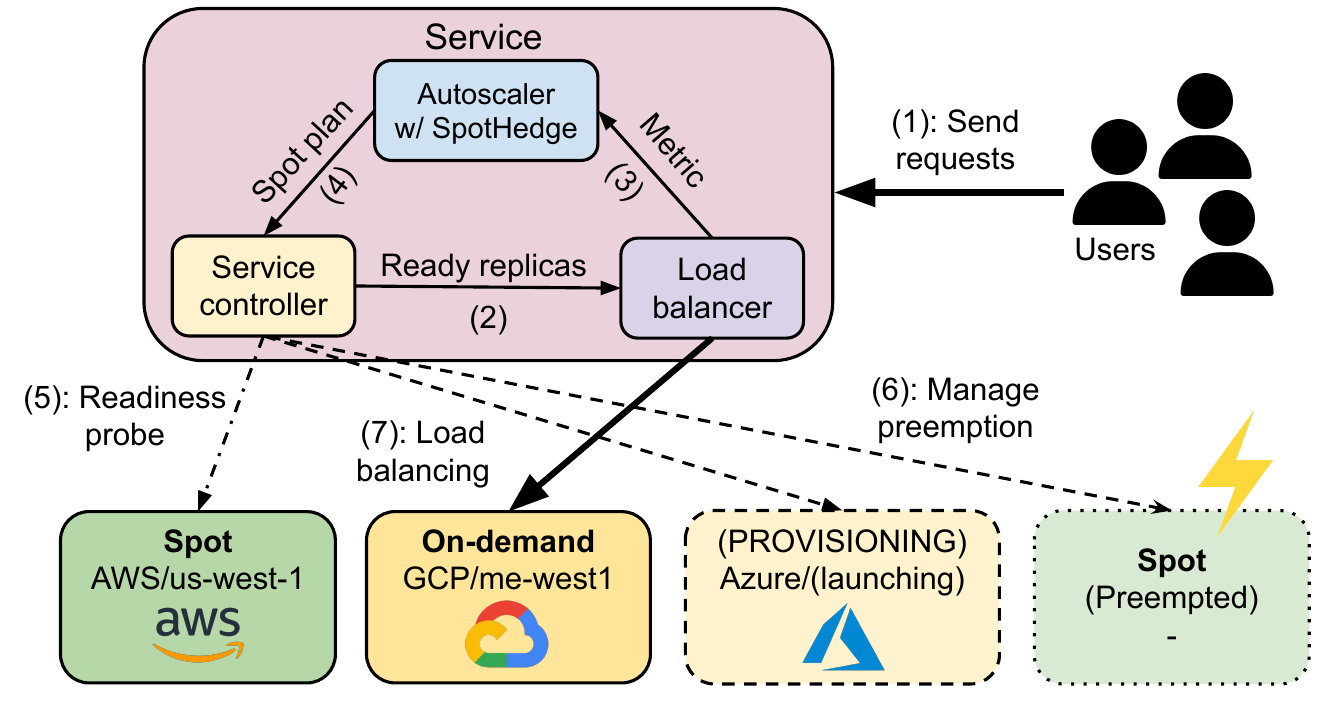}
    \caption{\textbf{An Overview of \sys (\secref{soda-service-controller}).} 
    Each service is composed of three components: a service controller, responsible for overseeing the provisioning and monitoring the health of replicas; an autoscaler with \policy to guide scaling decisions and decide spot-on-demand mixture and spot placement; and a load balancer for distributing traffic to replicas ready to handle requests.
    }
    \label{fig:skyserve-arch}
\end{figure}

We have implemented \sys (Figure~\ref{fig:skyserve-arch}), a prototype serving system leveraging \policy that manages a mixture of spot and on-demand replicas across multiple regions and clouds. It builds on an open-sourced multi-cloud system~\cite{yang2023skypilot} that provisions instances on public cloud providers. \sys adds both a serving system and our policy \policy with $\approx$ 7000 lines of Python code and supports common inference engines such as vLLM~\cite{kwon2023efficient}, TGI~\cite{tgi}, Triton~\cite{nvidia-triton}. 

\paragraph{Service Controller.} 
\label{soda-service-controller}
The service controller is responsible for overseeing the entire replica life cycle, including scaling up replicas in a zone, reducing extra on-demand replicas when there is a sufficient number of spot replicas or surplus replicas during periods of low request rates, and managing the preemptions of spot replicas or any arising errors (6). The service controller implements \textit{readiness\_probe}, a monitoring tool to periodically assess service status through either a standard health probe or an actual user-defined compute workload (5). It then forwards the ready replica information to the load balancer (2). The service controller periodically polls the cost information via cloud API used in Algorithm \ref{alg:dynamic-placement}.

\paragraph{Implementation of \policy.}
\label{soda-Autoscaler-policy}
\policy tries to maintain $N_{Tar}+N_{Extra}$ spot replicas, as described in~\secref{Spot-On-Demand-mixture}. These replicas may be in various states: provisioning, initializing, or ready for traffic handling. The system launches on-demand replicas when it does not have $N_{Tar}+N_{Extra}$ ready spot replicas. 
\policy schedules these on-demand replicas for termination when a sufficient number of spot replicas are ready to accept traffic. The system implements the spot placement policy outlined in~\secref{Spot-placement}.\label{Spot-hedge-impl-in-soda} It monitors the preemption activities of different zones and recommends a new active zone for launching new spot replicas. It then sends the spot plan, i.e. placement and fallback, to the service controller for execution in step (4).

\paragraph{Autoscaler.}
\label{soda-lb}
The system implements a load-based autoscaler. \policy decides the target number of replicas $N_{Tar}$ based on the target load \textit{per replica} $Q_{Tar}$.
The autoscaler keeps track of a configurable past time window (default to 1 minute) and calculates the average request rate to be $R_t$. The autoscaler proposes a candidate target $N_{Can} = \lceil \frac{R_t}{Q_{Tar}} \rceil$ and compare it with the current $N_{Tar}$. If $N_{Can}$ is consistently larger than the current $N_{Tar}$ for a certain amount of time (\eg 10 minutes), current $N_{Tar}$ is set to $N_{Can}$. Similarly, $N_{Tar}$ is decreased to $N_{Can}$ if $N_{Can}$ is consistently smaller than $N_{Tar}$. \sys additionally supports custom policies specified by the user, such as maintaining a minimum amount of on-demand capacity.

\paragraph{Load Balancer.}
The system load balancer distributes incoming traffic (1) and supports both round-robin and routing to replicas with the least number of ongoing requests (7). It also forward the metric measurements (\eg QPS) to the autoscaler for decision-making (3). It can be extended to route requests to replicas closer to the clients, prioritizing underloaded replicas.  
We elaborate more in \S\ref{discussion-future-work}. We leave these policies configurable to users. 

\paragraph{Application API} 
A service configuration (\eg Listing~\ref{code:service-app}) requires an endpoint, ranging from a conventional HTTP server to an AI model-based endpoint like LLMs or image generation models~\cite{rombach2021highresolution}. The example also specifies an endpoint for \texttt{readiness\_probe} (\eg a real compute workload specified at an endpoint \texttt{/v1/chat/completions}). 
The \texttt{replica\_policy} field specifies the configuration of SpotHedge, \eg the extra number of spot replicas to overprovision (\texttt{num\_overprovision}), and the desired QPS $Q_{Tar}$ for autoscaler (\texttt{target\_qps\_per\_replica}).

\begin{listing}[ht]
\scriptsize 
\yamlcode{auxiliary/service.yaml}
\caption{\textbf{\sys configuration for a LLM service.}}
\label{code:service-app}
\end{listing}

\paragraph{Support for distributed inference.}
\policy schedules multiple instances of the same replica to the same zone, while multiple replicas are placed across regions and clouds to minimize inter-region traffic.
Preemption of one spot instance will terminate the entire replica, and \policy can also adjust parallelization strategy over remaining spot instances similar to SpotServe~\cite{SpotServe}. 

\paragraph{Preemption handling.} \policy allows client-side retry upon preemption~\cite{openai-api}, or leveraging a proxy to retry on the client's behalf. The request will be aborted if the instance is preempted. A new copy of that request will be resent and reassigned to a ready replica. \policy can additionally leverage preemption warning by adjusting the on-demand and spot instance mixture upon receiving the warning and notifying the user application upon receiving cloud preemption warning to trigger user-defined preemption handlers. Unfortunately, as noted in \S\ref{spot-cpu-fail}, preemption warnings are best-effort and cannot resolve the availability issue.

\section{Evaluation}
\label{evaluation}
\label{experiment-methodology}

We conduct both end-to-end experiments (\secref{end-to-end-evaluation}) and experiments with simulated preemptions (\secref{microbenchmark}). In the former, we deploy \sys and compare it to several serving systems on the cloud with 
real-time preemptions; in the latter, we 
compare different policies based on real spot obtainability traces. 

\subsection{End-to-end Results with Real Preemptions on Cloud}
\label{end-to-end-evaluation}

We ran end-to-end experiments that lasted $\approx$22 hours in total and served 133k requests to compare \sys with several production and research systems with the total cost at \$4.1k. The experiments consist of two runs for each setup with \textit{all} compared systems running at the same time.

\paragraph{Baselines.} We compare with the following systems:
\begin{itemize}
\item \textbf{AWS Auto-scaling Group (ASG)~\cite{aws-autoscaling-group}}: ASG uses fixed node pools (\eg fixed percentage of spot replicas and on-demand replicas). 
We set the on-demand percentage to 10\% following its official example~\cite{aws-asg-official-guide}.
\item \textbf{MArk~\cite{zhang2019mark}}: an ML serving system focusing on spot CPU instances and using proactive autoscaling. We modify MArk to make it compatible with spot GPUs\footnote{MArk exclusively uses CPU instances when the request rate is low and employs burstable instances~\cite{aws-burstable} and Lambda~\cite{aws-lambda} which are not available for GPU instances. We modified MArk to use only GPU instances while keeping the remaining algorithms (\eg predictive autoscaling) the same.}. 
\item \textbf{AWS spot node pool (AWSSpot)~\cite{aws-autoscaling-group}}: A node pool that uses spot instances with autoscaling allocated over multiple zones of the same region.
\item \textbf{SpotServe~\cite{miao2023spotserve}}: SpotServe adapts parallelization strategies in response to preemption. SpotServe does not consider or implement instance provisioning, spot placement, scheduling, or autoscaling~\cite{SpotServe}. As such, we run SpotServe together with the above systems. 
\end{itemize}

\begin{figure}[t]
\begin{subfigure}{.48\linewidth}
    \centering
    \includegraphics[width=\linewidth]{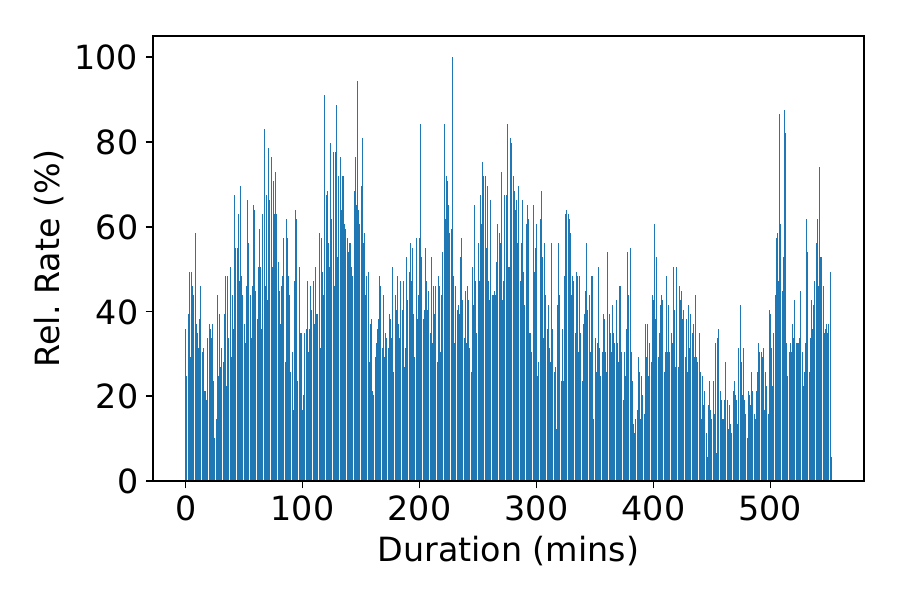}
    \caption{Request Arrival Pattern of the Arena Trace.}
    \label{fig:trace-timestamp}
\end{subfigure}
~
\begin{subfigure}{.48\linewidth}
    \centering
    \includegraphics[width=\linewidth]{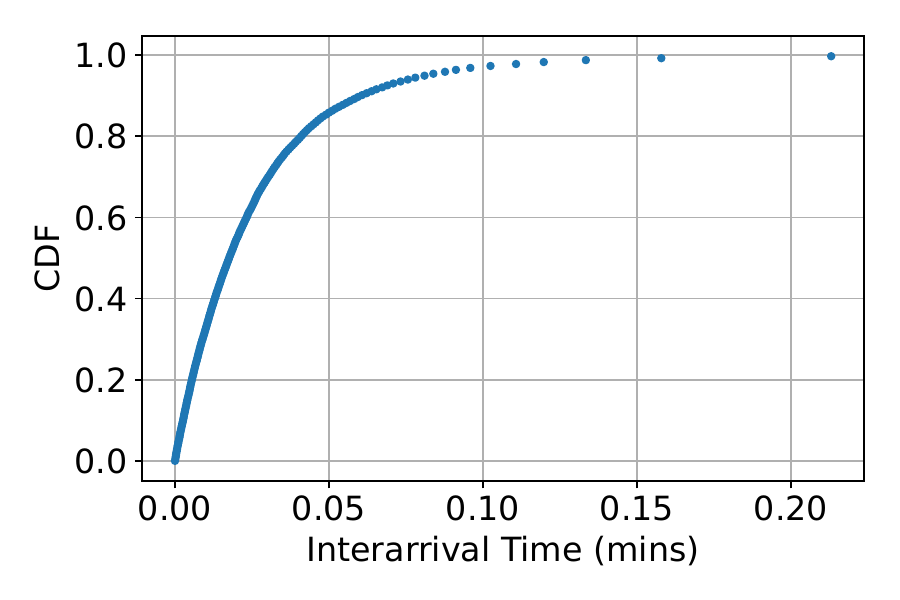}
    \caption{Request Interarrival Time Distribution of the Arena Trace.}
    \label{fig:trace-cdf}
\end{subfigure}
\caption{The request arrival pattern and the distribution of interarrival time of the Arena trace.}
\label{fig:arena-trace}
\end{figure}

\paragraph{Experiment Setup.}
We conduct an end-to-end evaluation on AWS, where baseline systems are launched \textit{concurrently} on the cloud and experience real-time preemptions, unavailability, and cold start delay for a fair comparison. 
We run two sets of experiments: (1) each replica runs on a \texttt{g5.48xlarge} instance (with 8 A10G GPUs) and consists of \text{Llama-2-70B}~\cite{touvron2023llama} using vLLM~\cite{kwon2023efficient}; (2) each replica runs on a \texttt{g4dn.12xlarge} instance (with 4 T4 GPUs) and consists of \text{OPT-6.7B} using SpotServe~\cite{miao2023spotserve}.
For workload, we use the inter-arrival time and query prompts from Arena (\secref{microbenchmark}), a real LLM serving workload from Chatbot Arena~\cite{chatbot-arena} with bursty traffic (Figure~\ref{fig:arena-trace}) and varying output lengths. 
Each serving system processes the same sequence of prompts, where each prompt is different and requires a different amount of processing time. Request timeouts are set to 100s for \text{Llama-2-70B} and 20s for \text{OPT-6.7B}, to account for the computation time of LLM requests. All requests that fail due to spot preemption will be retried by the client, with the failure time included in the overall end-to-end latency.
For \sys, we launch all replicas in the following regions: \texttt{us-east-2}, \texttt{us-west-2}, \texttt{eu-central-1}. We choose \texttt{us-west-2} for other baselines due to more quota, its popularity, and that its costs are lowest among regions. We conducted four experiments at different times of the week, and we categorized the experiment into two groups: \textit{Spot Available} 
(91\text{--}100\% spot obtainability\footnote{The spot obtainability (the percentage of time that a client can successfully launch an instance) is calculated with region \texttt{us-west-2}, where all of the baselines are launched.}) and \textit{Spot Volatile} (45\text{--}46\% spot obtainability). The cost is computed with the real-time price obtained via the cloud provider's API. Request failure rates are recorded by tracking request timeouts due to preemptions and queueing.

\begin{figure}[t]
\centering
\begin{subfigure}{.48\linewidth}
    \centering
    \includegraphics[width=\linewidth]{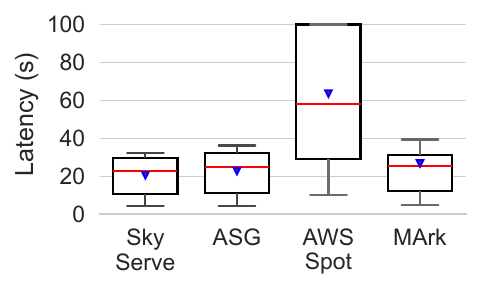}
    \caption{Latency, Spot Available.}
    \label{fig:e2e-latency-f4-f7-aggregated}
\end{subfigure}
~
\begin{subfigure}{.48\linewidth}
    \centering
    \includegraphics[width=\linewidth]{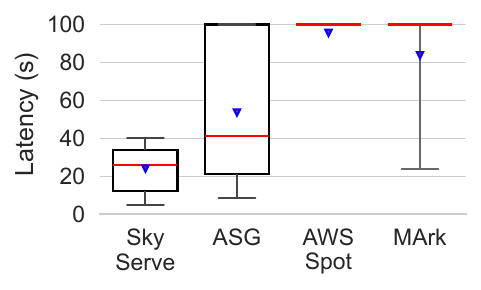}
    \caption{Latency, Spot Volatile.}
    \label{fig:e2e-latency-f6-f5-aggregated}
\end{subfigure}
\\
\begin{subfigure}{.48\linewidth}
    \centering    
    \includegraphics[width=\linewidth]{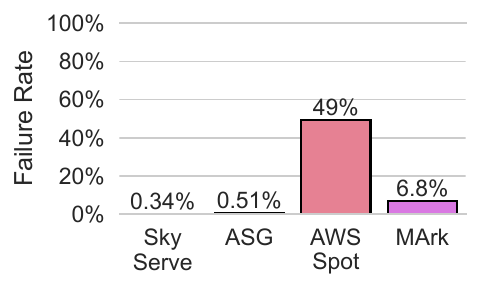}
    \caption{Failure Rate, Spot Available.}
    \label{fig:e2e-failure-rate-f4-f7-aggregated}
\end{subfigure}
~
\begin{subfigure}{.48\linewidth}
    \centering
    \includegraphics[width=\linewidth]{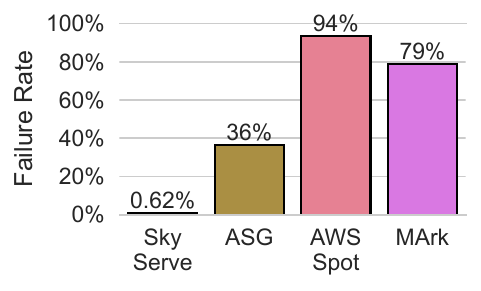}
    \caption{Failure Rate, Spot Volatile.}
    \label{fig:e2e-failure-rate-f6-f5-aggregated}
\end{subfigure}
\\
\begin{subfigure}{.48\linewidth}
    \centering    
    \includegraphics[width=\linewidth]{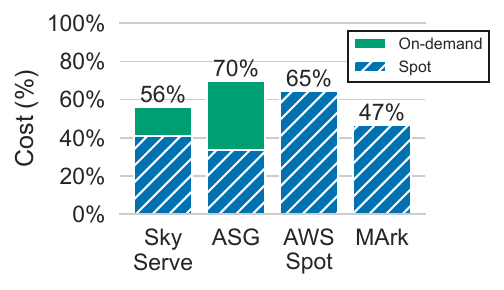}
    \caption{Cost, Spot Available.}
    \label{fig:e2e-cost-f4-f7-aggregated}
\end{subfigure}
~
\begin{subfigure}{.48\linewidth}
    \centering
    \includegraphics[width=\linewidth]{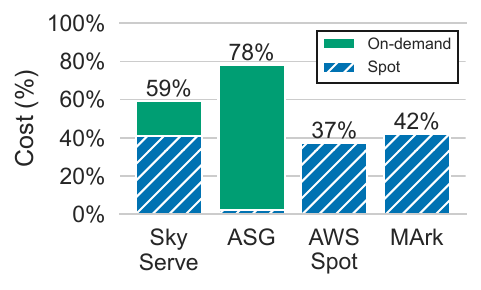}
    \caption{Cost, Spot Volatile.}
    \label{fig:e2e-cost-f5-f6-aggregated}
\end{subfigure}
\caption{
  \textbf{Service quality, failure rate, and cost.} We run \text{Llama-2-70B} on 8 A10G GPUs with vLLM.
Results are grouped by two scenarios (\secref{end-to-end-evaluation}): Spot Available and Spot Volatile.
For the box plot, the line marks the median, the box marks $25^{th}$ and $75^{th}$ percentiles, the whiskers show $10^{th}$ and $90^{th}$ percentiles, and the inverted triangle marks the mean. 
We show the cost breakdown of each group into either on-demand or spot. Costs are relative to using entirely on-demand instances (OD).} 
\label{fig:e2e-latency-and-failure-rate-aggregated}
\end{figure}

\begin{figure*}[t]
\centering
\begin{subfigure}{.35\linewidth}
    \centering
    \includegraphics[width=\linewidth]{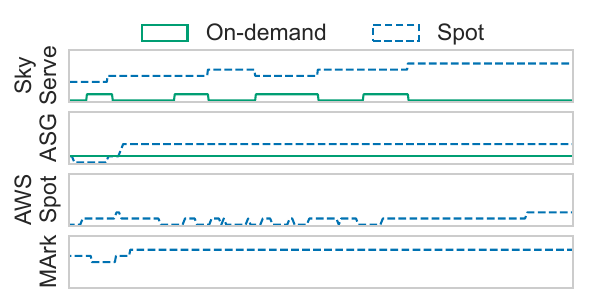}
    \caption{Spot Available.}
    \label{fig:e2e-num-node-to-time-f4-aggregated}
\end{subfigure}
~
\begin{subfigure}{.35\linewidth}
    \centering
    \includegraphics[width=\linewidth]{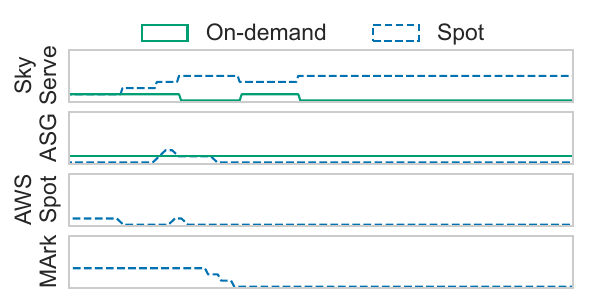}
    \caption{Spot Volatile.}
    \label{fig:e2e-num-node-to-time-f6-aggregated}
\end{subfigure}
\caption{We show the numbers of ready replicas launched by each system in two groups (Spot Available and Spot Volatile).}
\label{fig:e2e-num-node-to-time}
\end{figure*}

\paragraph{Service Quality and Failure Rate.} We run \text{Llama-2-70B} on 8 A10G GPUs (Figure~\ref{fig:e2e-latency-and-failure-rate-aggregated}). The number of spot and on-demand instances over time successfully provisioned is shown in Figure~\ref{fig:e2e-num-node-to-time}. Compared to ASG, \sys improves service quality (P50, P90, P99 latency) by
1.1\text{--}1.6$\times$, 1.1\text{--}2.5$\times$, 1.4\text{--}1.9$\times$ respectively. ASG maintains a single on-demand replica throughout the experiment. However, it encounters difficulties in acquiring additional spot replicas within one region due to spot unavailability. Consequently, ASG experiences a high failure rate of 36\% in the event of spot volatility. Increasing the number of always-on on-demand replicas in ASG can improve availability; however, it will significantly increase the cost of ASG deployment. Compared to AWSSpot, \sys largely improves P50, P90, and P99 latencies by 2.6\text{--}3.9$\times$, 2.5\text{--}3.1$\times$, and 1.9\text{--}2.7$\times$ as it leverages multiple regions to expand the available spot capacity and avoid highly-preempting zones. AWSSpot's single-region policy is unable to guarantee enough replicas, leading to a larger
49\text{--}94\% failure rate due to both service downtime and not enough replicas to cover client load, because AWSSpot's static even spread policy relaunches instances on highly-preempting zones and thus fails to get enough replicas. MArk has a failure rate of 6.8\text{--}79\%,
mainly due to periods of downtime where MArk cannot get any spot replicas (Figure~\ref{fig:e2e-num-node-to-time-f6-aggregated}). 
In comparison, \sys achieves high service availability with a low failure rate of 0.34\text{--}0.62\% across both groups.

\begin{figure}[t]
\centering
\begin{subfigure}{.47\linewidth}
    \centering    \includegraphics[width=\linewidth]{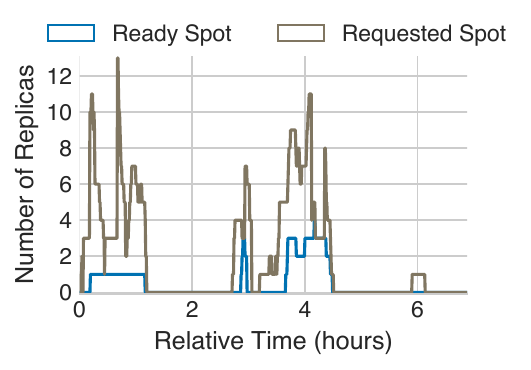}
    \caption{\textbf{MArk.}}
    \label{fig:mark-overprovision-to-10}
\end{subfigure}
~
\begin{subfigure}{.47\linewidth}
    \centering
    \includegraphics[width=\linewidth]{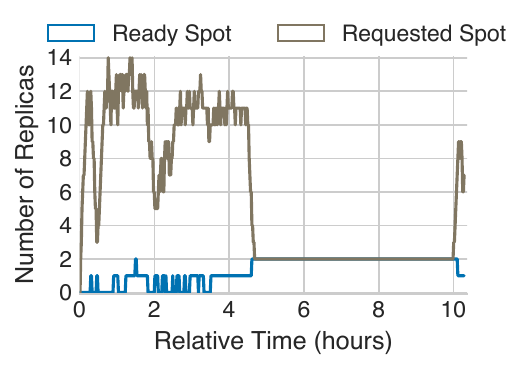}
    \caption{\textbf{AWSSpot.}}
    \label{fig:aws-spot-overprovision-to-14}
\end{subfigure}
\caption{MArk and AWSSpot over-request spot instances in case of spot unavailability, incurring extra cost. }
\label{fig:e2e-overrequest}
\end{figure}

\paragraph{Cost.}
Compared to ASG, \sys lowers cost by 20\text{--}24\%, as ASG maintains an on-demand replica even when spot instances are available. For ASG, in group Spot Volatile, the cost of on-demand replicas comprises 97\% of the total cost, since spot instances are less available. 
For AWSSpot, we observed a provision-then-preempt cycle in highly-preempting zones (\eg \texttt{us-west-2b}) when the spot obtainability is volatile with increased costs. 
MArk and AWSSpot over-request in the event of unavailability, and we observed up to 14 replicas in provisioning status (Figure~\ref{fig:aws-spot-overprovision-to-14}), likely because both systems target CPU instances and assume that replicas will quickly become ready after which provisioning replicas will be scaled down. 
Though AWSSpot uses entirely spot, \sys lowers cost by 20\% in group Spot Available. 
In group Spot Volatile, as both AWSSpot and MArk cannot provision enough spot replicas, they are cheaper than \sys (by 30\text{--}37\%). We observe that Dynamic Fallback adds a cost (\label{e2e-od-spot-price-ratio}26\text{--}31\% of the total cost) to \sys, though necessary to ensure availability. Compared to using only on-demand instances, \sys is 41\text{--}44\% cheaper.

\begin{figure}[t]
\centering
\begin{subfigure}{.48\linewidth}
    \centering
    \includegraphics[width=\linewidth]{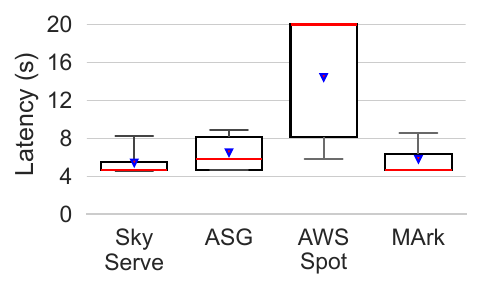}
    \caption{Latency, Spot Available.}
    \label{fig:spot-serve-latency}
\end{subfigure}
~
\begin{subfigure}{.48\linewidth}
    \centering
    \includegraphics[width=\linewidth]{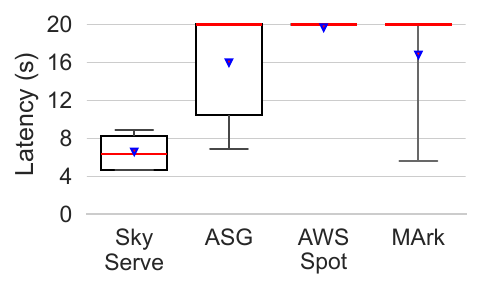}
    \caption{Latency, Spot Volatile.}
    \label{fig:spot-serve-latency}
\end{subfigure}
\\
\begin{subfigure}{.48\linewidth}
    \centering
    \includegraphics[width=\linewidth]{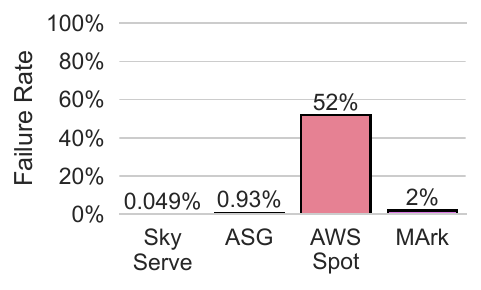}
    \caption{Failure Rate, Spot Available.}
    \label{fig:spot-serve-failure-rate}
\end{subfigure}
~
\begin{subfigure}{.48\linewidth}
    \centering
    \includegraphics[width=\linewidth]{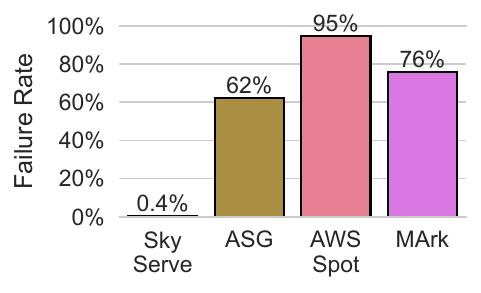}
    \caption{Failure Rate, Spot Volatile.}
    \label{fig:spot-serve-failure-rate}
\end{subfigure}
\\
\begin{subfigure}{.48\linewidth}
    \centering
    \includegraphics[width=\linewidth]{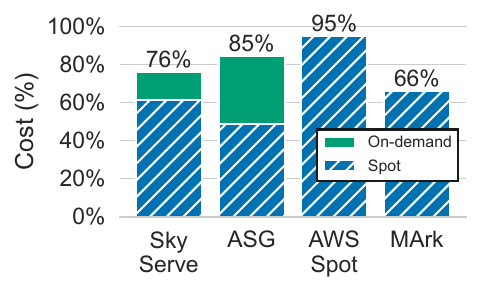}
    \caption{Cost, Spot Available.}
    \label{fig:spot-serve-cost-available}
\end{subfigure}
~
\begin{subfigure}{.48\linewidth}
    \centering
    \includegraphics[width=\linewidth]{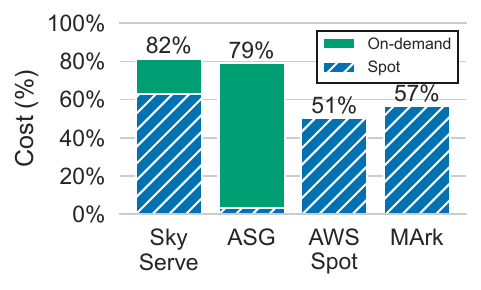}
    \caption{Cost, Spot Volatile.}
    \label{fig:spot-serve-cost-volatile}
\end{subfigure}
\caption{%
\textbf{Service quality, failure rate, and cost from running SpotServe~\cite{miao2023spotserve} (a baseline system) augmented with different systems}, serving \text{OPT-6.7B} on 4 T4 GPUs. SpotServe itself does not implement instance provisioning~\cite{SpotServe}. We find that \sys works well with SpotServe, saving cost for both Spot Available and Spot Volatile cases with good service latency and low failure rate. Although MArk or AWSSpot can have lower cost than \sys under Spot Volatile due to these systems provisioning fewer instances, using these systems leads to significantly higher failure rates and latency.}
\label{fig:spot-serve-other-system-trace}
\end{figure}

\paragraph{SpotServe~\cite{miao2023spotserve} running with baseline systems.} 
We compare \sys and other baseline systems on SpotServe, running OPT-6.7B model on 4 T4 GPUs. The latency, failure rates, and costs are presented in Figure~\ref{fig:spot-serve-other-system-trace}. 
Consistent with earlier figures, when spot instances are volatile, SpotHedge significantly reduces P50, P90, and P99 latency by 
3.1$\times$, 2.3$\times$, and 1.6$\times$
compared to baseline systems. The failure rate of \sys is only 0.05\text{--}0.4\%, significantly lower than other systems (52\text{--}95\%). This is because of \sys's ability to reduce preemptions by finding available zones across different regions and improve instance availability by dynamically adjusting spot and on-demand mixture. 
In Spot Available, MArk and \sys achieve similar P50 and P90 latency, but \sys improved P99 latency by 
2.2$\times$.
Compared to ASG, \sys improves service quality (P50, P90 and P99 latency) by 
1.2$\times$, 1.1$\times$ and 1.9$\times$.
AWSSpot experiences short periods of unavailability, primarily due to its static even spread policy, resulting in a 52\% failure rate. In terms of cost, when spot instances are available, \sys still manages to reduce cost by 10\text{--}20\% compared to ASG and AWSSpot, while achieving substantially better service and failure rates. MArk cannot launch enough instances and hence has a lower cost than \sys. 
ASG incurs a similar cost as \sys (3\% cheaper), as ASG keeps an on-demand instance at all times.

\paragraph{Discussion.}
We observe that on-demand instances are typically obtainable across regions and clouds, making it a reliable fallback. By running over multiple regions, \sys ensures that on-demand replicas can be quickly provisioned. In contrast, spot replicas can consistently be unobtainable in a single region. This observation is consistent with prior work~\cite{wu2024can}. We observed even an entire region (\texttt{us-west-2}) can run out of spot capacity. This shows the importance of spreading spot replicas across multiple regions. 
Though SpotServe adjusts parallelization among remaining instances upon preemption, naively using SpotServe in a single zone leads to poor service quality. Across different model sizes (\text{Llama-2-70B}  and \text{OPT-6.7B}), inference engines (vLLM and SpotServe), \sys is the only system with consistently low failure rates and request latency while achieving substantial cost savings compared to the traditional on-demand based serving setup. 

\begin{figure*}[t]
    \centering
    \begin{subfigure}{.3\linewidth}
        \centering
        \includegraphics[width=\linewidth]{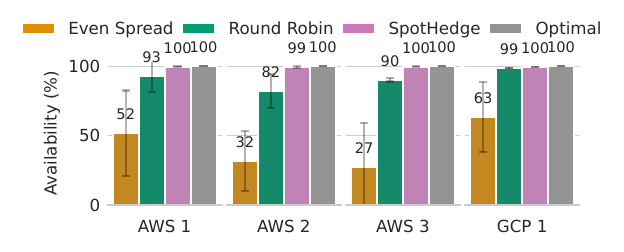}
        \caption{\textbf{Availability}}
        \label{fig:availability}
    \end{subfigure}   
    ~
    \begin{subfigure}{.29\linewidth}
        \centering
        \includegraphics[width=\linewidth]{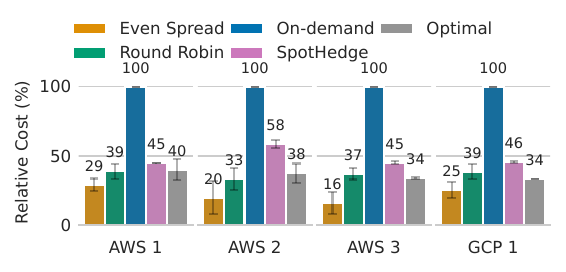}
        \caption{\textbf{Cost}}
        \label{fig:cost}
    \end{subfigure}
    ~
    \begin{subfigure}{.2\linewidth}
    \centering
    \includegraphics[width=\linewidth]{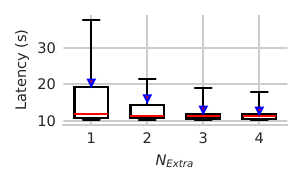}
    \caption{\textbf{Sensitivity to $N_{Extra}$}} 
    \label{fig:sensitivity-experiment-aws-2}
    \end{subfigure}
    ~
    \begin{subfigure}{.2\linewidth}
        \centering
        \includegraphics[width=\linewidth]{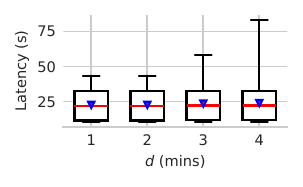}
        \caption{\textbf{Sensitivity to $d$}}
        \label{fig:sensitivity-experiment-aws-3}
    \end{subfigure}
    ~
    \caption{(a) and (b): Comparison of availability and cost across different spot traces and policies. We omit on-demand in the Availability experiment as it guarantees availability attainment. Cost is relative to $N_{Tar}$ on-demand replicas. (c), (d) Sensitivity to the number of spot replicas to overprovision ($N_{Extra}$) and cold start delay ($d$) under Poisson workload. The same trend generalizes to other workloads and traces. }
    \label{fig:combined}
\end{figure*}

\subsection{Results with Simulated Preemptions from Real Traces}
\label{microbenchmark}

We run the following workloads individually: 

\begin{itemize}
\item \textbf{Poisson Distribution (Poisson).} We construct synthetic workloads with Poisson distribution of request arrival ($\lambda = 0.15$). 
\item \textbf{Arena.} We use a real LLM serving workload from Chatbot Arena~\cite{chatbot-arena} with load fluctuations, bursty traffic (Figure~\ref{fig:arena-trace}), and dynamic request execution time. 

\item \textbf{Microsoft Azure Function (MAF).} MAF~\cite{shahrad2020serverless} was collected from Azure serverless function invocations over two weeks, and has been used for ML serving research~\cite{li2023alpaserve, ishakian2018serving, bhattacharjee2019barista, miao2023spotserve}. 
\end{itemize}

Following prior work~\cite{miao2023spotserve, wu2024can}, we benchmark policy performance based on \textit{real} spot obtainability traces and workload traces. Instead of experiencing real-time preemptions on the cloud (\S\ref{end-to-end-evaluation}), spot preemptions are instead injected based on the collected spot obtainability traces.
We compare \policy with the following  policies, similar to prior work~\cite{yang2023snape}:

\begin{itemize}
\item \textbf{Even Spread}: Evenly spreading spot replicas across zones of different regions, similar to~\cite{aws-autoscaling-group, zhang2019mark, ray-serve}. 
\item \textbf{Round Robin}: Launch spot replicas to zones of different regions,  
round-robin. 
\item \textbf{Optimal}: Omniscient policy based on ILP (\secref{omniscient}). 
\end{itemize}

\paragraph{Spot datasets.} We use spot traces from~\cite{wu2024can}. These traces were previously collected by launching spot GPUs and experiencing preemptions and unobtainability on the cloud. Each timestamp records the number of preemptions experienced when maintaining the desired number of spot instances.

\begin{itemize}
    \item \textbf{AWS 1:} A 2-week trace for 4 p3.2xlarge in 3 zones. 
    \item \textbf{AWS 2:} A 3-week trace for 16 p3.2xlarge in 3 zones. 
    \item \textbf{AWS 3:} A 2-month trace for p3.2xlarge in 9 zones. 
    \item \textbf{GCP 1:} A 3-day trace for 4 a2-ultragpu-4g in 6 zones.  
\end{itemize}

\begin{figure*}[t]
\centering
\begin{subfigure}{.33\linewidth}
    \centering    \includegraphics[width=\linewidth]{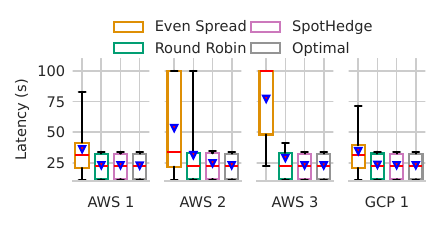}
    \caption{\textbf{Poisson}}
    \label{fig:latency-box-Poi}
\end{subfigure}
~
\begin{subfigure}{.33\linewidth}
    \centering    \includegraphics[width=\linewidth]{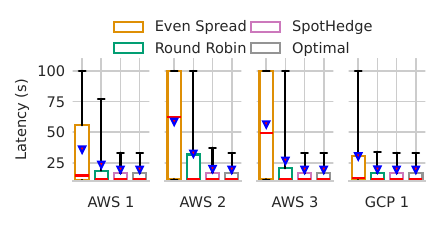}
    \caption{\textbf{Arena}}
    \label{fig:latency-box-Arena}
\end{subfigure}
~
\begin{subfigure}{.33\linewidth}
    \centering\includegraphics[width=\linewidth]{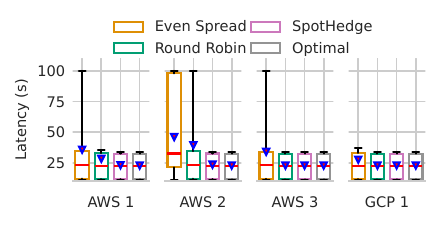}
    \caption{\textbf{MAF}}
    \label{fig:latency-box-MAF}
\end{subfigure}
\caption{\textbf{Service Latency} across four spot traces and three workloads. 
}
\label{fig:latency-box}
\end{figure*}

\paragraph{Availability.}
We evaluate the service availability of the service deployment (percentage of time a certain number of instances is ready to serve requests) achieved by different policies (Figure~\ref{fig:availability}). Even Spread achieves 27\text{--}63\% availability, whereas Round Robin achieves 82\text{--}99\% availability. When spot replicas become unavailable, Even Spread and Round Robin cannot ensure that we can quickly find and launch at least $N_{Tar}$ ready spot replicas. For example, in AWS 2, spot replicas experience unavailability across all zones. In these cases, Even Spread and Round Robin will suffer unavailability. Round Robin incurs more unavailability when there are highly-preempting zones, because it will retry in those zones. In contrast, \policy achieves high availability ($99\text{--}100\%$) for three reasons. First, Dynamic Fallback ensures that \policy uses on-demand replicas when spot replicas become unavailable. Second, it is unlikely to lose many spot replicas before an on-demand fallback replica is ready, thanks to overprovisioned spot replicas as a buffer. Third, by tracking highly-preempting zones, \policy can minimize the likelihood of preemptions by avoiding these zones. 

\paragraph{Cost.}
\policy reduces cost by 42\text{--}55\% (Figure~\ref{fig:cost}) compared to using entirely on-demand replicas. 
The cost for Even Spread (16\text{--}29\% of on-demand) and Round Robin (33\text{--}39\% of on-demand) is lower, due to more spot preemptions and less spot capacity. This is consistent with the end-to-end experiments, where these policies cannot launch enough replicas, suffering from worse service quality with lower costs. 
Omniscient policy (\S\ref{omniscient}) does not overprovision and leverages future information to minimize cost. \policy achieves 5\%\text{--}20\% relative cost difference to Optimal, while achieving comparable resource availability.

\paragraph{Service Quality.}

We report request latency 
in Fig~\ref{fig:latency-box}. 
A static policy (\eg Even Spread) experiences frequent preemptions, resulting in fewer ready replicas than $N_{Tar}$ and worse service quality. \policy achieves 1.1\text{--}3.0$\times$, 1.0\text{--}1.8$\times$ reduction in average latency compared to Even Spread and Round Robin, and within 5\% to the Optimal.

\paragraph{Sensitivity Experiments.}
We show (Fig.~\ref{fig:sensitivity-experiment-aws-2}, Fig.~\ref{fig:sensitivity-experiment-aws-3}) sensitivity of latency to the number of spot replicas to overprovision ($N_{Extra}$) and cold start delay ($d$). We find that a small value for $N_{Extra}$ works well in practice. A larger cold start delay moderately increases the tail latency.

\section{Future Work}
\label{discussion-future-work}

\paragraph{Advanced load balancing policy.} For requests that require real-time interactivity and short Time-to-first-token (TTFT), \policy can be extended to dynamically route these requests to replicas in the same zone as the client, as well as monitor the replica load and only direct requests to a remote zone if local replicas are overloaded.
\policy can also keep an on-demand node pool in the client zone for these requests. This ensures that most requests can still be served from the closest replica to the client, while the requests that would otherwise cause overload can be routed to another replica in a remote region. That said, 
we expect inter-region latency to be largely outweighed by computation time for most requests (\S\ref{Spot-multi-region}).
 
\paragraph{Support heterogeneous accelerators.} GPUs have different performance-to-cost trade-offs. 
Expensive GPUs tend to have better performance albeit being less obtainable. While \sys supports specifying a set of GPUs, it can be extended to adopt a more intelligent policy to leverage accelerator heterogeneity. For example, when the spot instance for a higher-end GPU is unobtainable, one might switch to a spot instance of a cheaper, lower-end GPU instead. 

\section{Related Work}
\label{related-works}

\paragraph{Spot instances for non-serving workloads.} 
Spot instances have drawn interest from both industry~\cite{spot-io, yang2023snape} and academia~\cite{song2012optimal,yi2011monetary, wu2024can}. Prior work explored using spot instances for maintaining resource availability~\cite{yang2023snape}, web services~\cite{mazzucco2011achieving}, in-memory storage~\cite{xu2016blending},  batch~\cite{wolski2017probabilistic,domanal2018efficient} or interactive jobs~\cite{ambati2019optimizing}, HPC~\cite{taifi2011spotmpi}, analysis tasks~\cite{lee2017deepspotcloud}, or elastic services~\cite{harlap2018tributary}. ML training on spot instances has also been extensively studied~\cite{shang2023spotdnn, harlap2017proteus, thorpe2023bamboo, athlur2022varuna, wagenlander2020spotnik, yang2021scheduling}. For example, Bamboo provides resilience for DNN training on preemptible instances~\cite{thorpe2023bamboo}. Varuna~\cite{athlur2022varuna} trains DNN networks on spot instances with commodity networking. However,  \textit{serving} has received relatively less attention and requires high resource availability and low latency.

\paragraph{Spot instances for serving workloads.} 
As mentioned, SpotServe addresses a different set of problems from \policy: preemption-tolerant model parallelization across \textit{multiple} spot instances. Other works that incorporate spot instances are MArk~\cite{zhang2019mark}, Cocktail~\cite{gunasekaran2022cocktail}, and Tributary~\cite{harlap2018tributary}. However, they focus on serving small ML models using spot CPUs. In~\secref{spot-gpu-different-cpu}, we showed that spot GPUs are more likely to experience preemptions than spot CPUs, and systems that target spot CPUs do not perform well for serving large AI models (\secref{end-to-end-evaluation}). Further, prior work only considers allocating instances in a single region or zone. \sys instead provisions GPUs across multiple regions and clouds to improve service availability. Snape~\cite{yang2023snape} uses spot CPU instances for model serving by introducing an RL framework that uses node-level and cluster-level aggregate capacity to predict spot obtainability. However, this information is Azure-internal. 

\paragraph{Popular model serving solutions.}
Many proprietary solutions host AI services. SageMaker~\cite{aws-sagemaker} allows users to deploy ML models on AWS. Google Vertex AI~\cite{google-vertex-ai}  helps deploy GenAI models into applications on Google. However, these systems are not open-sourced and focus on non-preemptible instances. Ray Serve~\cite{ray-serve} is a model-serving library that supports both spot and on-demand instances. Similar to AWS Autoscaling Group~\cite{aws-autoscaling-group}, Ray Serve only supports predefined node pools and redirects traffic to on-demand nodes upon spot preemptions. GKE, a popular managed Kubernetes service, supports both spot and on-demand instances but does not support automatically scaling down on-demand insatnces once spot obtainability comes back~\cite{aws-gke-od-spot-mix}, or support multi-region. 

\section{Conclusions}

Spot GPUs are economically appealing to serve AI workloads, but they have not been widely considered viable for serving due to unpredictable preemptions and long cold start delays.
We introduce \policy to serve AI workloads on a mixture of spot and on-demand GPUs across regions and clouds. \policy diversifies spot replica placements across regions and clouds, overprovisions spot replicas to mitigate preemptions, and proactively uses on-demand replicas when spot replicas are less available. 
We implement \sys that leverages \policy and provides a unified interface to host AI services on mixtures of spot and on-demand replicas. Through evaluations on real AI workloads, \sys saves cost by 43\% on average while achieving similar resource availability compared to on-demand deployments, and improves P50, P90, and P99 latency by 2.3$\times$, 2.1$\times$, 2.1$\times$ on average respectively compared to other systems.

\section*{Acknowledgement}

We thank our shepherd, John Wilkes, and anonymous
EuroSys reviewers for their valuable comments and insightful
feedback. This work is in part supported by gifts from
Accenture, AMD, Anyscale, Google, IBM, Intel, Microsoft, Mohamed Bin Zayed University of Artificial Intelligence,
Samsung SDS, Uber, and VMware.

\bibliographystyle{ACM-Reference-Format}
\bibliography{bib/paper}

\clearpage
\appendix
\section{Appendix}
\label{appendix}

This Appendix contains the code and instructions to reproduce the experiments and figures in this paper. We open-sourced SkyServe system to stimulate research in this area: \bluelink{https://github.com/skypilot-org/skypilot}. We also open-source all collected spot preemption traces in this repository: \bluelink{https://github.com/MaoZiming/spothedge_ae}.

\subsection{Reproducing Experiment Results Locally}
We provide all raw data for our experiments and scripts to reproduce all figures locally. 

\paragraph{Preparations}

Clone our repository and install dependencies:

\begin{lstlisting}[basicstyle=\ttfamily\small, breaklines=true]
git clone https://github.com/MaoZiming/spothedge_ae
conda create --name spot-hedge-ae python=3.10
conda activate spot-hedge-ae
pip install -r requirements.txt
\end{lstlisting}

\paragraph{Reproduce End-to-End Experiment Figures}

We provide raw data for our experiments and scripts to reproduce all figures locally. To reproduce the end-to-end experiment figures:

\begin{lstlisting}[basicstyle=\ttfamily\small, breaklines=true]
cd path/to/ae/repo
cd e2e/plots
unzip spot-hedge-ae-raw-data.zip
# Figure 6
python3 draw-misc.py
# Figure 9(a-d), Figure 13(a-d)
python3 draw-latency.py
# Figure 9(e-f), Figure 10, Figure 12, Figure 13(e-f)
python3 draw-cost-and-trace.py
\end{lstlisting}

\paragraph{Reproduce Microbenchmark Experiment Figures}

To reproduce the microbenchmark experiment figures:

\begin{lstlisting}[basicstyle=\ttfamily\small, breaklines=true]
cd path/to/ae/repo
cd plots
./plot_figures.sh
\end{lstlisting}

\subsection{Running SkyServe on the Cloud}

\paragraph{Install SkyServe}

Install SkyServe from source following this guide: \bluelink{https://docs.skypilot.co/en/latest/getting-started/installation.html}. To use \policy, switching to the \texttt{spot-hedge-new} branch and enable at least one cloud. We use AWS for example.

\begin{lstlisting}[basicstyle=\ttfamily\small, breaklines=true]
conda create -y -n sky python=3.10
conda activate sky
git clone https://github.com/skypilot-org/skypilot.git
cd skypilot
git switch spot-hedge-new
pip install -e ".[aws]"
sky check aws # And follow the instructions.
\end{lstlisting}

\paragraph{Spinning up an AI service}

Here are some example commands to run an AI service. This launches an OpenAI API Server for Llama2-70b-chat-hf. Notice that a HuggingFace access token is required at line 40 of \texttt{e2e/spot\_hedge.yaml}. For more information, checkout the SkyServe system: \bluelink{https://docs.skypilot.co/en/latest/serving/sky-serve.html}.

\begin{lstlisting}[basicstyle=\ttfamily\small, breaklines=true]
cd path/to/ae/repo
# Add your HuggingFace token at line 40
sky serve up e2e/spot_hedge.yaml -n spot-hedge
\end{lstlisting}

Once running, use the following command to monitor service status:

\begin{lstlisting}[basicstyle=\ttfamily\small, breaklines=true]
watch -n10 sky serve status spot-hedge
\end{lstlisting}

SpotHedge automatically handles preemption recovery and auto-scaling. Once ready, the endpoint URL will be available as an OpenAI API Server. Send a test query command:

\begin{lstlisting}[basicstyle=\ttfamily\small, breaklines=true]
ENDPOINT=$(sky serve status spot-hedge --endpoint)
curl $ENDPOINT/v1/models
curl $ENDPOINT/v1/chat/completions -X POST -H "Content-Type: application/json" -d '{
  "model": "meta-llama/Llama-2-70b-chat-hf",
  "messages": [
    {
      "role": "user",
      "content": "Hello! What is your name?"
    }
  ]
}'
\end{lstlisting}

\paragraph{Launch clients for the end-to-end experiments}

The configuration of clients is located in \texttt{e2e/client/client.yaml}. To run it, replace the host and port with the actual service host and port.

\begin{lstlisting}[basicstyle=\ttfamily\small, breaklines=true]
# Get the service host and port
sky serve status spot-hedge --endpoint
# Replace it in e2e/client/client.yaml
# Start the client
sky launch e2e/client/client.yaml
\end{lstlisting}

\paragraph{Running Microbenchmark}

YAML files for microbenchmarks are in \texttt{eval/run\_eval.yaml}. To reproduce it, run:

\begin{lstlisting}[basicstyle=\ttfamily\small, breaklines=true]
sky launch eval/run_eval.yaml
\end{lstlisting}

\end{document}